\documentclass[10pt,preprint]{aastex}

\newcommand{\Sersic}{S\'ersic}

\newcommand{\avg}[1]{{\langle{#1}\rangle}}

\def\simless{\mathbin{\lower 3pt\hbox
	{$\,\rlap{\raise 5pt\hbox{$\char'074$}}\mathchar"7218\,$}}} 
\def\simgreat{\mathbin{\lower 3pt\hbox
	{$\,\rlap{\raise 5pt\hbox{$\char'076$}}\mathchar"7218\,$}}} 

\newcommand{\petroratio}{{{\mathcal{R}}_P}}
\newcommand{\petroradius}{{{r}_P}}

\newcommand{\petroratiolim}{{{\mathcal{R}}_{P,\mathrm{lim}}}}
\newcommand{\band}[2]{\ensuremath{^{#1}\!{#2}}}
\newcommand{\Vmax}{\ensuremath{V_\mmax}}
\newcommand{\mmax}{\ensuremath{\mathrm{max}}}
\newcommand{\mmin}{\ensuremath{\mathrm{min}}}
\newcommand{\minmax}{\ensuremath{\mathrm{\left\{^{min}_{max}\right\}}}}
\newcommand{\fixMr}{\ensuremath{M_\fixr}}
\newcommand{\fixr}{\ensuremath{\fixmag{r}}}
\newcommand{\fixredshift}{0.1}
\newcommand{\fixmag}[1]{\ensuremath{{^{\fixredshift}\!{#1}}}}

\newcounter{thefigs}
\newcommand{\fignum}{\arabic{thefigs}}

\newcounter{thetabs}

\newcounter{address}

\slugcomment{Submitted to \apj}

\shortauthors{Blanton {\it et al.} (2002)}
\shorttitle{Galaxy Properties}

\begin{document}
 

\title{The Broad-band Optical Properties of Galaxies with Redshifts
$0.0 < z < 0.2$\altaffilmark{\ref{SDSS}}}



\author{
Michael R. Blanton\altaffilmark{\ref{NYU}},
David W. Hogg\altaffilmark{\ref{NYU}},
Neta A. Bahcall\altaffilmark{\ref{Princeton}},
Ivan K.~Baldry\altaffilmark{\ref{JHU}},
J. Brinkmann\altaffilmark{\ref{APO}},
Istv\'an Csabai\altaffilmark{\ref{Eotvos},\ref{JHU}},
Daniel Eisenstein\altaffilmark{\ref{Steward}},
Masataka Fukugita\altaffilmark{\ref{CosmicRay}},
James E. Gunn\altaffilmark{\ref{Princeton}},
\v{Z}eljko Ivezi\'{c}\altaffilmark{\ref{Princeton}},
D. Q.~Lamb\altaffilmark{\ref{Chicago}},
Robert H. Lupton\altaffilmark{\ref{Princeton}},
Jon Loveday\altaffilmark{\ref{Sussex}},
Jeffrey A. Munn\altaffilmark{\ref{USNO}},
R. C. Nichol\altaffilmark{\ref{CarnegieMellon}},
Sadanori Okamura\altaffilmark{\ref{Tokyo}},
David J. Schlegel\altaffilmark{\ref{Princeton}},
Kazuhiro Shimasaku\altaffilmark{\ref{Tokyo}},
Michael A. Strauss\altaffilmark{\ref{Princeton}},
Michael S. Vogeley\altaffilmark{\ref{Drexel}} and
David H. Weinberg\altaffilmark{\ref{Ohio}}
}

\altaffiltext{1}{Based on observations obtained with the
Sloan Digital Sky Survey\label{SDSS}} 
\setcounter{address}{2}
\altaffiltext{\theaddress}{
\stepcounter{address}
Center for Cosmology and Particle Physics, Department of Physics, New
York University, 4 Washington Place, New
York, NY 10003
\label{NYU}}
\altaffiltext{\theaddress}{
\stepcounter{address}
Princeton University Observatory, Princeton,
NJ 08544
\label{Princeton}}
\altaffiltext{\theaddress}{
\stepcounter{address}
Department of Physics and Astronomy, The Johns Hopkins University,
Baltimore, MD 21218
\label{JHU}}
\altaffiltext{\theaddress}{
\stepcounter{address}
Apache Point Observatory,
2001 Apache Point Road,
P.O. Box 59, Sunspot, NM 88349-0059
\label{APO}}
\altaffiltext{\theaddress}{\stepcounter{address}
Department of Physics, E\"{o}tv\"{o}s University,
Budapest, Pf.\ 32, Hungary, H-1518
\label{Eotvos}}
\altaffiltext{\theaddress}{
\stepcounter{address}
Steward Observatory, 933 N. Cherry Ave., Tucson, AZ 85721
\label{Steward}}
\altaffiltext{\theaddress}{
\stepcounter{address}
Institute for Cosmic Ray Research, University of
Tokyo, Midori, Tanashi, Tokyo 188-8502, Japan
\label{CosmicRay}}
\altaffiltext{\theaddress}{
\stepcounter{address}
University of Chicago, Astronomy \&
Astrophysics Center, 5640 S. Ellis Ave., Chicago, IL 60637
\label{Chicago}}
\altaffiltext{\theaddress}{
\stepcounter{address}
Sussex Astronomy Centre,
University of Sussex,
Falmer, Brighton BN1 9QJ, UK
\label{Sussex}}
\altaffiltext{\theaddress}{
\stepcounter{address}
U.S. Naval Observatory,
3450 Massachusetts Ave., NW,
Washington, DC  20392-5420
\label{USNO}}
\altaffiltext{\theaddress}{
\stepcounter{address}
Department of Physics, Carnegie Mellon University, 
5000 Forbes Avenue, Pittsburgh, PA 15213-3890 
\label{CarnegieMellon}}
\altaffiltext{\theaddress}{
\stepcounter{address}
Department of Astronomy and Research Center for 
the Early Universe,
School of Science, University of Tokyo,
Tokyo 113-0033, Japan
\label{Tokyo}}
\altaffiltext{\theaddress}{
\stepcounter{address}
Department of Physics, Drexel University, Philadelphia, PA 19104
\label{Drexel}}
\altaffiltext{\theaddress}{
\stepcounter{address}
Ohio State University,
Department of Astronomy,
Columbus, OH 43210
\label{Ohio}}

\begin{abstract}
Using photometry and spectroscopy of 144,609 galaxies from the Sloan
Digital Sky Survey, we present bivariate distributions of pairs of
seven galaxy properties: four optical colors, surface brightness,
radial profile shape as measured by the \Sersic\ index, and absolute
magnitude. In addition, we present the dependence of local galaxy
density (smoothed on 8 $h^{-1}$ Mpc scales) on all of these
properties.  Several classic, well-known relations among galaxy
properties are evident at extremely high signal-to-noise ratio: the
color--color relations of galaxies, the color--magnitude relations,
the magnitude--surface brightness relation, and the dependence of
density on color and absolute magnitude. We show that most of the
$i$-band luminosity density in the universe is in the absolute
magnitude and surface brightness ranges used: $-23.5 <
M_{\band{0.1}{i}} < -17.0~\mathrm{mag}$ and $17 < \mu_{\band{0.1}{i}}
< 24~\mathrm{mag}$ in $1~\mathrm{arcsec^{2}}$ (the notation
$\band{z}{b}$ represents the $b$ band shifted blueward by a factor
$(1+z)$). Some of the relationships between parameters, in particular
the color--magnitude relations, show stronger correlations for
exponential galaxies and concentrated galaxies taken separately than
for all galaxies taken together. We provide a simple set of fits of
the dependence of galaxy properties on luminosity for these two sets
of galaxies.
\end{abstract}


%
%

\section{Motivation}
\label{motivation}

There are strong correlations among the measurable physical properties
of galaxies.  The classification of galaxies along the visual morphological
sequence described by \cite{hubble36a} correlates well with the
dominance of their central bulge, their surface brightnesses, and
their colors. These properties also correlate with other properties,
such as metallicity, emission-line strength, luminosity in visual
bands, neutral gas content, and the winding angle of the spiral
structure (for a review, see \citealt{roberts94a}). The surface
brightnesses of giant galaxies classified morphologically as
elliptical are known to be strongly correlated with their sizes
(\citealt{kormendy77a}). Galaxy colors (at least of morphologically
elliptical galaxies) are known to be strongly correlated with galaxy
luminosity (\citealt{baum59,faber73a,visvanathan77,terlevich01a}). The
gravitational mass of a galaxy is closely related to the luminosity
and other galaxy properties. These galaxy relations manifest
themselves in the Tully-Fisher relation for spiral galaxies
(\citealt{tully77a,burstein97a}) and the Fundamental Plane for
morphologically elliptical galaxies
(\citealt{faber76a,djorgovski87,dressler87,burstein97a});
\citet{strauss95a} review these dynamical relations. Furthermore, the
environment of a galaxy is related to its type, in the sense that
early-type galaxies are found in denser regions than late-type
galaxies, as first noted by \citet{hubble36a} and as found by numerous
subsequent investigators (\citealt{oemler74a,dressler80a,davis76a,
giovanelli86a,santiago92a,guzzo97a,hashimoto99a,blanton00a}). In
short, different physical properties of galaxies are closely related
to each other.

In order to understand theoretically how galaxies formed and acquired
their current properties, it is first necessary to characterize the
observed distribution of galaxy properties in a comprehensive manner.
Such a characterization is a primary goal of the Sloan Digital Sky
Survey (SDSS; \citealt{york00a}).  The SDSS has already created the
largest sample to date (144,609 galaxies) of luminous galaxies with
well-measured photometric and spectroscopic properties. As a first
step in understanding the joint distribution of galaxy properties, we
present here a study of the number density and luminosity density
distributions of galaxy colors, profiles, luminosities, surface
brightnesses, and local densities. Our purpose is to measure
properties related to quantities that cosmological gasdynamical
simulations (\citealt{nagamine01a}; \citealt{pearce01a};
\citealt{steinmetz02a}) or semi-analytic models
(\citealt{somerville01a}; \citealt{mathis02a}) can predict or will
soon be able to predict, such as galaxy ages, sizes, stellar masses,
and degrees of concentration.


The paper is organized as follows. Section \ref{data} briefly
describes the spectroscopic sample of the SDSS. Section
\ref{properties} describes how we measure the set of properties
studied here.  Section \ref{results} shows the number density and
luminosity density distributions of these properties of galaxies, and
provides simple fits for the dependence on luminosity of the properties
of galaxies with nearly exponential radial profiles and those with
more concentrated radial profiles. Section \ref{conclusion}
summarizes and outlines future lines of research.

\section{Data}
\label{data}

\subsection{Description of the Survey}

The SDSS (\citealt{york00a}) is producing imaging and spectroscopic
surveys over $\pi$ steradians in the Northern Galactic Cap. A
dedicated wide-field 2.5m telescope (Siegmund et al., in preparation)
at Apache Point Observatory, Sunspot, New Mexico, images the sky in
five bands between 3000 and 10000 \AA\ ($u$, $g$, $r$, $i$, $z$;
\citealt{smith02a}) using a drift-scanning, mosaic CCD camera
(\citealt{gunn98a}), detecting objects to a flux limit of $r\sim 22.5$
mags. The photometric quality of the observations are tracked using an
automatic photometricity monitor (\citealt{hogg01a}). An automatic
image-processing pipeline known as {\tt photo} (Lupton et al. in
preparation) processes the images and produces a catalog. The
astrometric calibration is also performed by an automatic pipeline
(\citealt{pier02a}).  One of the goals is to spectroscopically observe
900,000 galaxies, (down to $r_{\mathrm{lim}}\approx 17.77$ mags),
100,000 Luminous Red Galaxies (\citealt{eisenstein01a}), and 100,000
quasars (\citealt{fan99a}, \citealt{richards02a}).  This spectroscopic
follow up uses two digital spectrographs on the same telescope as the
imaging camera. Many of the details of the galaxy survey are described
in the galaxy target selection paper (\citealt{strauss02a}). Other
aspects of the survey are described in the Early Data Release paper
(EDR; \citealt{stoughton02a}). The survey has begun in earnest, and to
date has obtained about 30\% of its intended data.

The SDSS images are reduced and catalogs are produced by the SDSS
pipeline {\tt photo}, which measures the sky background and the seeing
conditions, and detects and measures objects. The photometry used here
for the bulk of these objects was the same as that used when the
objects were targeted. However, for those objects which were in the
EDR photometric catalog, we used the better calibrations and
photometry from the EDR. The versions of the SDSS pipeline {\tt photo}
used for the reductions of the data used here ranged from {\tt v5\_0}
to {\tt v5\_2}. The treatment of relatively small galaxies, which
account for most of our sample, did not change substantially
throughout these versions.  The astrometric calibration is also
performed by an automatic pipeline which obtains absolute positions to
better than 0.1 arcsec (\citealt{pier02a}).  

The magnitudes are calibrated to a standard star network
(\citealt{smith02a}) approximately in the AB system. There are small
differences between the system output by the SDSS pipelines and a true
AB system, amounting to $\Delta m = -0.042, 0.036, 0.015, 0.013,
-0.002$ in the $u$, $g$, $r$, $i$, and $z$ bands. Because these were
discovered at a relatively late date in the preparation of this
manuscript, we have not self-consistently included these shifts in our
results (that is, by recalculating $K$-corrections based on the
revised colors). Instead we have applied them {\it a posteriori} to
our results in the $\band{0.1}{u}$, $\band{0.1}{g}$, $\band{0.1}{r}$,
$\band{0.1}{i}$, and $\band{0.1}{z}$ bands.

Object fluxes are determined several different ways by {\tt photo}, as
described in \citet{stoughton02a}. The primary measure of flux used
for galaxies is the SDSS Petrosian magnitude, a modified version of
the quantity proposed by \citet{petrosian76a}. The essential feature
of Petrosian magnitudes is that in the absence of seeing they measure
a constant fraction of a galaxy's light regardless of distance (or
size).  They are described in greater detail below (and also by
\citet{blanton01a} and \citet{strauss02a}). As measures of the galaxy
profile shape, {\tt photo} also calculates the radius containing half
the Petrosian flux ($r_{50}$) and that containing 90\% of the
Petrosian flux ($r_{90}$). 

As described in \citet{strauss02a}, the bulk of the fibers in the SDSS
are allocated to three samples: the Main Sample galaxies, the Luminous
Red Galaxies, and quasars. Here we will concern ourselves solely with the
Main Sample galaxies. There are three main criteria for their
selection:
\begin{eqnarray}
r_{\mathrm{PSF}} - r_{\mathrm{model}} &>& s_{\mathrm{limit}} \cr
r_{\mathrm{petro}} &<& r_{\mathrm{limit}}, \mathrm{~and}\cr
\mu_{50} &<& \mu_{50,\mathrm{limit}}.
\end{eqnarray}
Here, $r_{\mathrm{PSF}}$ is an estimate of the magnitude using the
local PSF as a weighted aperture, $r_{\mathrm{model}}$ is an estimate
of the magnitude using the better of a de~Vaucouleurs and an
exponential fit to the image (accounting for seeing),
$r_{\mathrm{petro}}$ is a modified form of the magnitude described by
\citet{petrosian76a}, and $\mu_{50}$ is the half-light surface
brightness, defined as the average surface brightness within the
radius which contains half of the Petrosian flux. All of these
quantities are reddening-corrected using the dust maps of
\citet{schlegel98a}. In practice, the values of the target selection
parameters vary across the survey in a well-understood way, but for
the bulk of the area, they are: $s_{\mathrm{limit}}=0.3$,
$r_{\mathrm{limit}}=17.77$ mag, and $\mu_{50,\mathrm{limit}}=24.5$ mag
in 1 arcsec$^2$.  Objects near the spectroscopic flux limit are nearly
five magnitudes brighter than the photometric limit; that is, the
fluxes are measured at signal-to-noise ratio of a few hundred.

Fibers are assigned to a set of circular tiles with a field of view
$1.49$ deg in radius by an automatic tiling pipeline
(\citealt{blanton02a}).  The targets are observed by a multi-fiber
spectrograph capable of taking 640 spectra simultaneously (48 of the
fibers are, in normal SDSS operations, devoted to sky fibers and
spectrophotometric standards). An automatic data processing pipeline
{\tt spec2d} wavelength- and flux-calibrates the spectra and outputs a
one-dimensional spectrum for each object (Schlegel, et al., in
preparation). The resulting spectra have a resolution of $R\sim 2000$,
cover roughly 3800 \AA\ to 9000 \AA, and typically have
signal-to-noise per pixel around 10. This spectrum is then analyzed by
another pipeline ({\tt spectro1d}; SubbaRao, et al., in preparation)
which classifies the spectrum and determines the redshift of the
object. The redshifts used in the current analysis are determined
independently using a separate pipeline originally designed
specifically for bright stars({\tt specBS}; Schlegel, et al., in
preparation) whose results are nearly identical for the main galaxy
sample to the ``official'' SDSS redshift determination (over 99\% of
the objects obtain nearly identical redshifts in each pipeline).

As of April 2002, the SDSS had imaged and targeted 2,873 deg$^2$ of
sky and taken spectra of approximately 350,000 objects over $\sim
2,000$ deg$^2$ of that area. The completeness over this area is
roughly 91\%; most of the missing galaxies (7\% of the total) are lost
due to the fact that two fibers on the same tile cannot be placed more
closely than 55$''$, about 1\% are lost because they are not able to
be assigned a fiber, and about 1\% cannot have a redshift determined.
From these data, we created a well-defined sample for calculating
large-scale structure and galaxy property statistics, known as
Large-Scale Structure {\tt sample10}. {\tt sample10} consists of all
of the photometry for all of the targets over that area (as extracted
from the internal SDSS operational database), all of the spectroscopic
results (as output from {\tt specBS}), and, most significantly, a
description of the angular window function of the survey and the flux
and surface brightness limits used for galaxies in each area of the
sky. For most of the area, the same version of the analysis software
used to create the target list was used in this sample. However, for
the area covered by the Early Data Release (EDR;
\citealt{stoughton02a}) we used the version of the analysis software
used for that data release, since it was substantially better than the
early versions of the software used to target that area. For {\tt
photo}, the most important piece of analysis software run on the data,
the versions used for the photometry range from {\tt v5\_0} to {\tt
v5\_2}.  The region covered by this sample is similar to, but not
exactly, the region which will be released in the SDSS Data Release 1
(DR1), scheduled for January 1, 2003 (which will use {\tt photo
v5\_3}, a newer version of the software which among other things
improves the handling of large galaxies).

\begin{figure}
\figurenum{\fignum}
\plotone{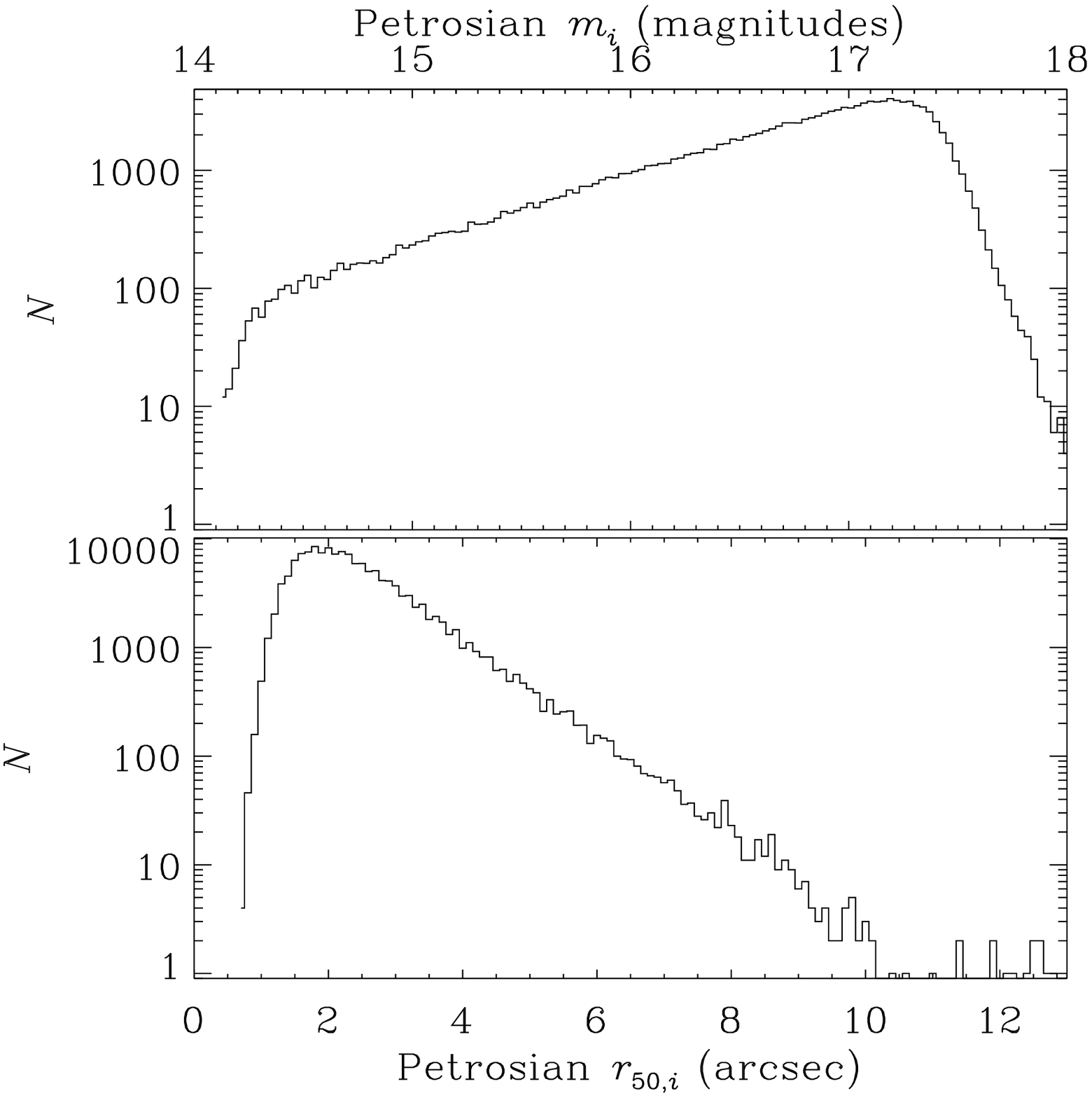}
\caption{\label{emg_imagr50} Distribution of reddening-corrected
$i$-band Petrosian magnitude (top panel) and the observed $i$-band
Petrosian half -light radius. The roll-offs at faint and bright
magnitudes are due to the explicit conditions in our sample that $14.5
< r < 17.77$ (which are accounted for in our $1/V_{\mathrm{max}}$
weighting).}
\end{figure}

For the purposes of this paper, we further restrict {\tt sample10} to
galaxies in the redshift range $0.02 < z < 0.22$, the absolute
magnitude range $-23.5 < M_{\band{0.1}{i}} < -17.0$ (assuming a Hubble
constant $H_0 = 100h$ km$^{-1}$ s$^{-1}$ Mpc$^{-1}$ and $h=1$), and to
galaxies with $r > 14.5$.  To calculate the distance modulus and the
comoving volume we assume a model with $\Omega_0 =0.3$ and
$\Omega_\Lambda =0.7$. These cuts restrict our sample to 144,609
galaxies. Figure \ref{emg_imagr50} shows the distributions of $i$-band
apparent magnitude (corrected for Galactic extinction) and Petrosian
half-light radius for this sample of galaxies. The roll-off at the
brightest and faintest magnitudes occurs because we have selected the
sample in the $r$-band.

\section{Galaxy Properties}
\label{properties}

Here we describe the set of galaxy properties whose joint distribution
we intend to measure, and how those properties are measured by the
SDSS. These properties are colors,
luminosities, profile shapes, surface brightnesses, and local
densities.

\subsection{Galaxy Colors and Luminosities}

To measure the galaxy fluxes, we rely on the SDSS Petrosian
magnitudes. The essential feature of Petrosian magnitudes is that in
the absence of seeing they measure a constant fraction of a galaxy's
light regardless of distance (or size).  More specifically, we define
the ``Petrosian ratio'' $\petroratio$ at a radius $r$ from the center
of an object to be the ratio of the local surface brightness averaged
over an annulus at $r$ to the mean surface brightness within $r$:
\begin{equation}
\label{petroratio}
\petroratio (r)\equiv \frac{\left.
\int_{\alpha_{\mathrm{lo}} r}^{\alpha_{\mathrm{hi}} r} dr' 2\pi r'
I(r') \right/ \left[\pi(\alpha_{\mathrm{hi}}^2 -
\alpha_{\mathrm{lo}}^2) r^2\right]}{\left.
\int_0^r dr' 2\pi r'
I(r') \right/ [\pi r^2]},
\end{equation}
where $I(r)$ is the azimuthally averaged surface brightness profile
and $\alpha_{\mathrm{lo}}<1$, $\alpha_{\mathrm{hi}}>1$ define the
annulus.  The SDSS has adopted $\alpha_{\mathrm{lo}}=0.8$ and
$\alpha_{\mathrm{hi}}=1.25$.  The Petrosian radius $\petroradius$ is
the radius at which the $\petroratio$ falls below a threshhold value
$\petroratiolim$, set to 0.2 for the SDSS. The Petrosian flux is
defined as the flux within a circular aperture with a radius equal to
$N_P \petroradius$, where $N_P = 2$ for the SDSS.  Petrosian
magnitudes are described in greater detail by \citet{blanton01a} and
\citet{strauss02a}.  While SDSS Petrosian magnitudes contain over 99\%
of the flux within an exponential profile, they contain only around
80\% of the flux within a de~Vaucouleurs profile in the absence of
seeing; the fraction is higher in the presence of seeing, as shown by
\citet{blanton01a}. For the SDSS, $\petroradius$ is defined in the
$r$-band and used to set the aperture for all the bands, such that the
colors represent a measure of spectral energy distribution of the
galaxy over a well-defined region of the galaxy.

Here we constrain the spectral energy distribution of the galaxies
using the broad-band colors between the $u$, $g$, $r$, $i$, and $z$
Petrosian magnitudes of the SDSS. A given set of observer-frame galaxy
colors does not correspond uniquely to a particular SED, so we cannot
infer with certainty the colors in any other frame. However, we can
use our knowledge of the form of galaxy SEDs based on other data to
make an attempt in order to interpolate between the observed-frame
colors to a different frame by fitting a reasonable model SED to the
galaxy colors.  To do so, we use the routines in {\tt kcorrect
v1\_11}, described by \citet{blanton02b}, which uses a similar method
to the method of \citet{csabai00a} for calculating photometric
redshifts. This method is demonstrably better than using the spectra
(which suffer from the aperture biases associated with small-fiber
spectroscopy, the inaccuracy of spectrophotometry, and the fact that
we do not take spectra in the $u$ and $z$ bands; {\it cf.}
\citealt{kochanek00a}) and better than simply interpolating the
magnitudes between bands (which does not account well for the 4000
\AA\ break).

\begin{figure}
\figurenum{\fignum}
\plotone{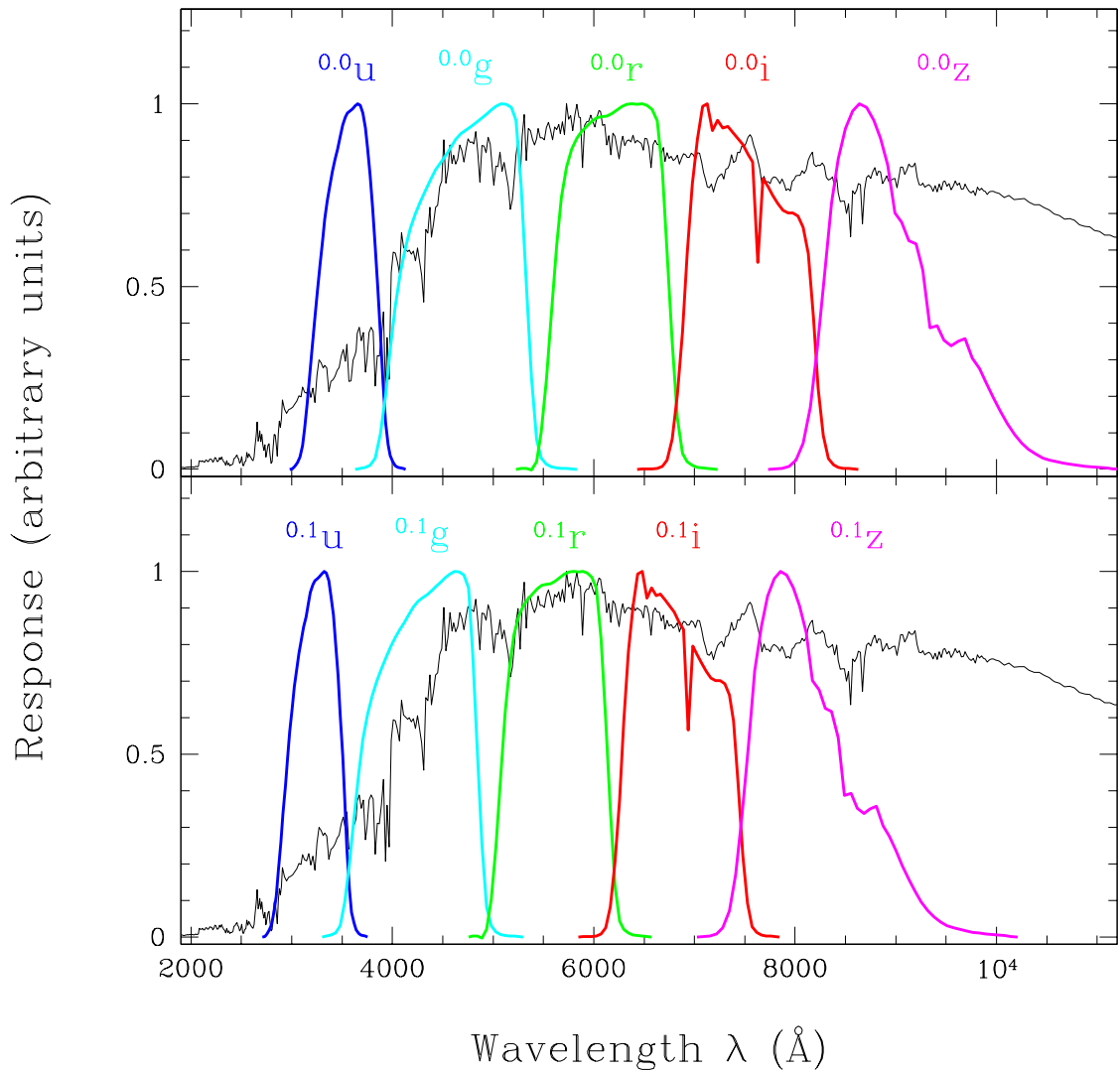}
\caption{\label{response} Demonstration of the differences between the
unshifted SDSS filter system (\band{0.0}{u}, \band{0.0}{g},
\band{0.0}{r}, \band{0.0}{i}, \band{0.0}{z}) in the top panel and the
SDSS filter system shifted by $0.1$ (\band{0.1}{u}, \band{0.1}{g},
\band{0.1}{r}, \band{0.1}{i}, \band{0.1}{z}) in the bottom
panel. Shown for comparison is a 4 Gyr-old instantaneous burst
population from an update of the \citet{bruzual93a} stellar population
synthesis models. The $K$-corrections between the magnitudes of a
galaxy in the unshifted SDSS system observed at redshift $z=0.1$ and
the magnitudes of that galaxy in the $0.1$-shifted SDSS system
observed at redshift $z=0$ are independent of the galaxy's spectral
energy distribution (and for AB magnitudes are equal to $-2.5
\log_{10} (1+0.1)$ for all bands; \citealt{blanton02b}). This
independence on spectral type makes the $0.1$-shifted system a more
appropriate system in which to express SDSS results, for which the
median redshift is near redshift $z=0.1$.}
\end{figure}

The essence of the method is that each set of observed magnitudes in
the five bands is fit by a linear combination of four templates. The
coefficients in the linear combination, and the templates themselves,
are optimized to reproduce the observed magnitudes.  For each galaxy,
we now have a model for the SED.  We use this SED to calculate the
correction from the observed band to fixed frame bandpasses with
shapes corresponding the the restframe coverage of galaxies at
$z=0.1$, the median redshift of the sample.  The usual choice is to
correct the colors to the $z=0$ frame; however, this choice is not
optimal, because most galaxies in our sample are not at $z=0$.  Figure
\ref{response} shows the responses of the unshifted SDSS and the SDSS
system shifted to $0.1$.  We refer to bands in this system as
$\band{0.1}{u}$, $\band{0.1}{g}$, $\band{0.1}{r}$, $\band{0.1}{i}$,
and $\band{0.1}{z}$, following the nomenclature of
\citet{blanton02b}. We note in passing that the $K$-correction for a
galaxy {\it at} redshift $z=0.1$ to this system is $-2.5 \log_{10}
(1.1)$ for all bands independent of galaxy SED.

\begin{figure}
\figurenum{\fignum}
\plotone{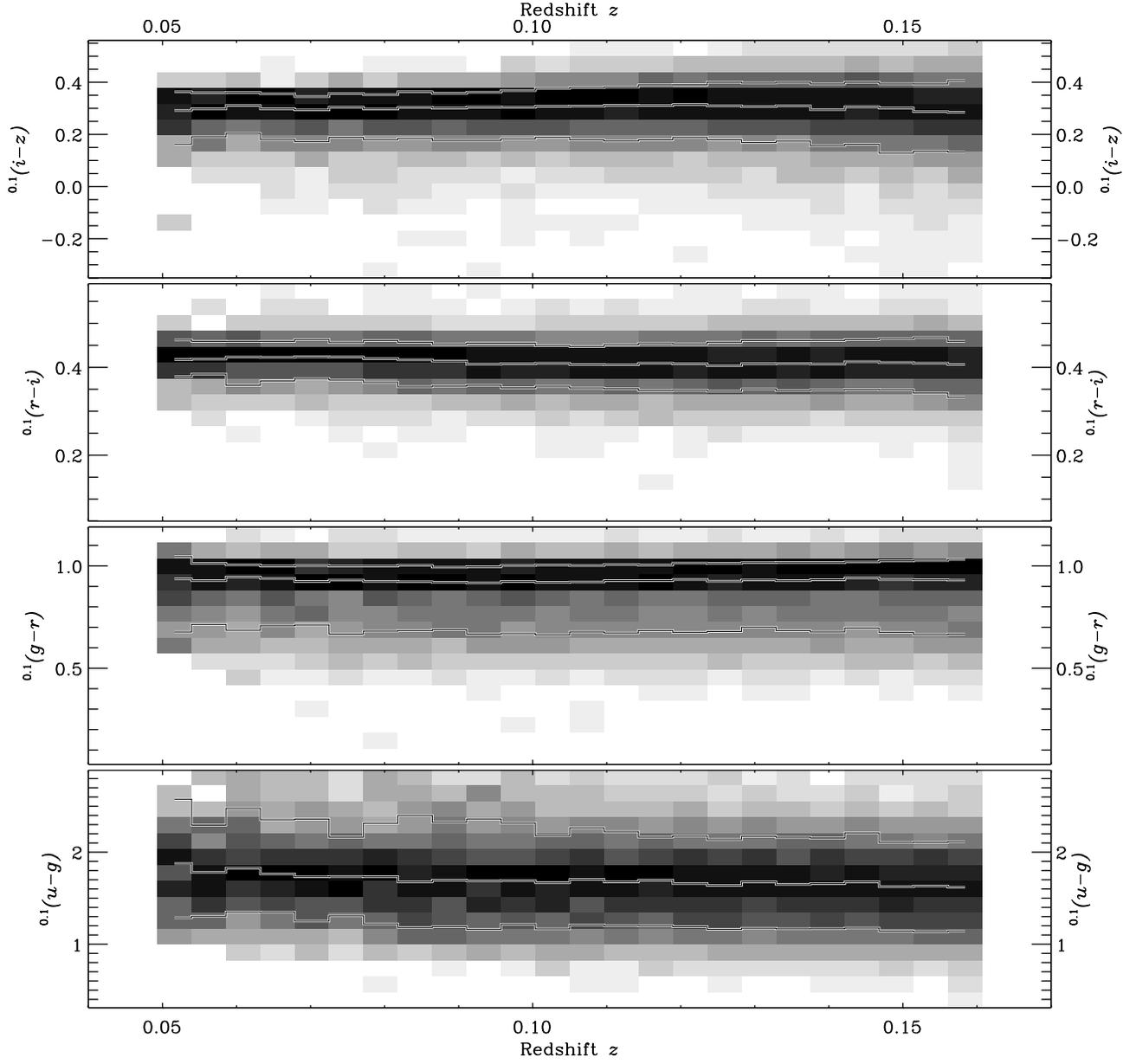}
\caption{\label{emg_colors} Distribution of fixed-frame colors at
$z=0.1$ as a function of redshift for a volume limited sample in the
ranges $0.05 < z < 0.16$ and $-22.5 < M_{\band{0.1}{r}} < -21.5$. The
greyscale is the conditional probability distribution of color on
redshift, which is to say that each column is normalized
separately. The lines indicate the 10\%, 50\% and 90\% quantiles in
each column. That the color distributions do not strongly evolve
indicates that the $K$-corrections are reasonable. Some of the
evolution apparent in the plot --- for example the fact that the
median $\band{0.1}{(u-g)}$ color becomes bluer with cosmic time ---
may well reflect real galaxy evolution.}
\end{figure}

As a test of the validity of the colors, Figure \ref{emg_colors} shows
the distribution of galaxy colors as a function of redshift in a
volume-limited sample in the range $-22.5 < M_{\band{0.1}{i}} < -21.5$
and $0.05 < z < 0.16$. The greyscale is the conditional distribution
of color on redshift. The middle line in each panel is the median, the
outer lines are the 10\% and 90\% quantiles. The distribution of
galaxy colors is nearly constant with redshift, indicating that the
$K$-corrections are relatively sensible. Note that we expect some
change in these distributions given that galaxies evolve; in fact
there is a hint of such evolution in the $\band{0.1}{u-g}$ plot.

We use these estimated Petrosian $\band{0.1}{(u-g)}$,
$\band{0.1}{(g-r)}$, $\band{0.1}{(r-i)}$ and $\band{0.1}{(i-z)}$
colors as our measures of the SED shape for each galaxy. As a measure
of luminosity, we use the $i$-band flux converted to
$\band{0.1}{i}$-band luminosity luminosity using the $K$-correction
and the standard cosmological formulae for the distance modulus,
assuming $\Omega_m = 0.3$ and $\Omega_\Lambda = 0.7$
(\citealt{hogg99cosm}).

\subsection{Galaxy Profile Shape and Surface Brightness}

One measure of the morphology of galaxies is the radial dependence of
their surface brightness. For example, if one expresses this radial
dependence in the form (suggested by \citealt{sersic68a}):
\begin{equation}
\label{sersic}
I(r) = A \exp\left[ - (r/r_0)^{1/n} \right],
\end{equation}
galaxies with nearly exponential profiles (\Sersic\ index $n\sim 1$)
tend to be considered ``late-type,'' while galaxies with nearly
de~Vaucouleurs profiles (\Sersic\ index $n \sim 4$) tend to be
considered ``early-type.'' For this reason, we use measures of the
azimuthally-averaged radial profiles of the galaxies in our
analysis. 

\begin{figure}
\figurenum{\fignum}
\plotone{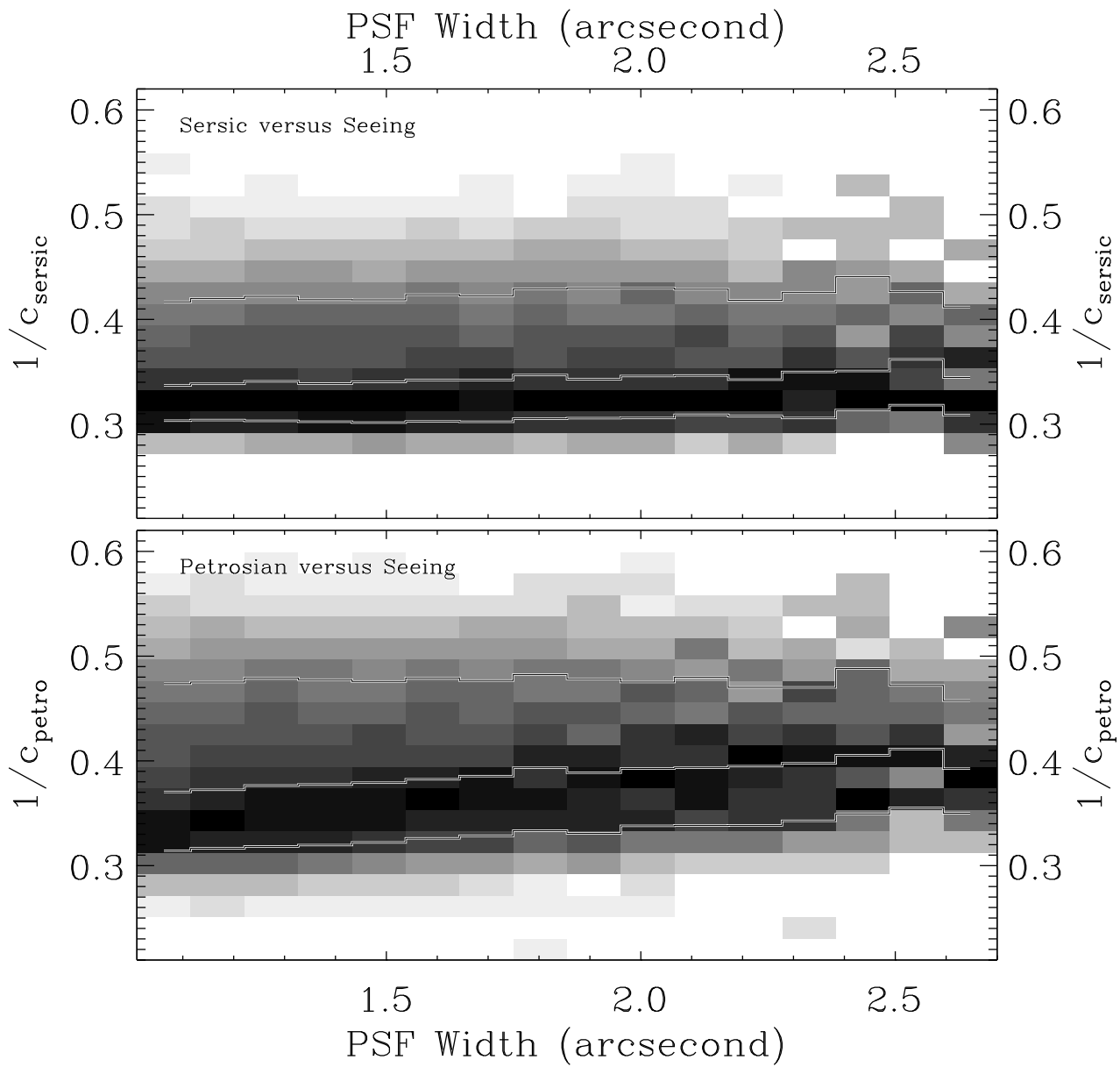}
\caption{\label{emg_seeing} Dependence of the inverse concentration
index on seeing for a volume limited sample in the ranges $0.04 < z <
0.11$ and $-21 < M_{\band{0.1}{i}} < -20$. Each panel shows as a grey
scale the conditional distribution of the $1/c\equiv r_{50}/r_{90}$
given the width of the seeing in arcseconds. (Because this is a
conditional distribution, the greyscale does not indicate the amount
of data which exists at a given seeing). As in Figure
\ref{emg_colors}, the lines indicate the 10\%, 50\% and 90\% quantiles
in each column. The bottom panel shows the measured Petrosian $1/c$ as
a function of seeing. There is a clear dependence of measured
concentration on seeing; the median changes from around 0.37 to 0.42
from the best to worst seeing, respectively. Remember that any
dependence on seeing also implies a dependence on apparent size,
introducing a complicated interdependence of measured concentration
with the actual concentration, the intrinsic size of the galaxy, the
distance, and the seeing conditions.  The top panel shows the
Petrosian $1/c$ that would be measured from the \Sersic\ profile fit
in the absence of seeing. This quantity shows almost no dependence on
seeing conditions, indicating that this quantity eliminates, to some
degree, the dependence of measured profile shape on observing
conditions and galaxy size.}
\end{figure}

Although \citet{strateva01a} and \citet{shimasaku01a} have shown that
the Petrosian inverse concentration index $1/c\equiv r_{50}/r_{90}$ is
reasonably well correlated with eyeball morphological classifications
for nearby, large galaxies, this parameter is too sensitive to the
effects of seeing to be of great use for the bulk of the SDSS Main
Sample. The top panel of Figure \ref{emg_seeing} shows the dependence
of inverse concentration parameter on seeing for a volume-limited
subsample, limited to galaxies with $-21 < M_{\band{0.1}{i}} < -20$
and $0.04 < z < 0.11$, demonstrating this dependence. This seeing
dependence means that the same galaxy, observed under different
conditions, can have considerably different measured
concentrations. Even more importantly, it implies that the same
galaxy, observed at different {\it distances}, will have different
measured concentrations.

{\tt photo} fits two-dimensional models to the galaxy images which
account for the effects of seeing, but they are limited to the cases
of $n=1$ and $n=4$, and in addition (because they were designed to
study faint galaxy colors) for the versions of {\tt photo} used here
(before {\tt v5\_3}) they only performed the fit out to a relatively
small radius compared to the size of an SDSS spectroscopic target
galaxy. Thus, for our work here we have decided to fit the parameters
in Equation (\ref{sersic}) to the radial profile of each galaxy
(accounting for the effects of seeing).

In order to do so, we find the parameters $A$, $r_0$, and $n$ in
Equation (\ref{sersic}) which minimize $\chi^2$ with respect to the
observed $i$-band radial profile and errors (as expressed by the {\tt
profMean} and {\tt profErr} parameters in the SDSS catalog). When
performing the fit, we convolve with the axisymmetric double Gaussian
fit to the PSF determined by the SDSS photometric pipeline (Lupton et
al., in preparation). The resulting distribution of $\chi^2$ per
degree of freedom peaks at approximately unity, but has a much larger
tail than expected for gaussian errors; this suggests that a more
detailed model (either for the seeing or for galaxy profiles) which
better fits the data is possible to develop. After performing the fit,
we can calculate the Petrosian inverse concentration parameter
$r_{50}/r_{90}$ based on the \Sersic\ index $n$. The top panel of
Figure \ref{emg_seeing} shows this inverse concentration parameter as
a function of seeing for the volume-limited subsample. Our measure of
galaxy profile shape is less seeing-dependent than the raw estimate
$r_{50}/r_{90}$.  We have fit \Sersic\ profiles to the profiles of all
bands; the results of the $g$, $r$, $i$, and $z$ profiles are all
fairly consistent, while the $u$ band results tend to be much less
concentrated than the others.

\begin{figure}
\figurenum{\fignum}
\plotone{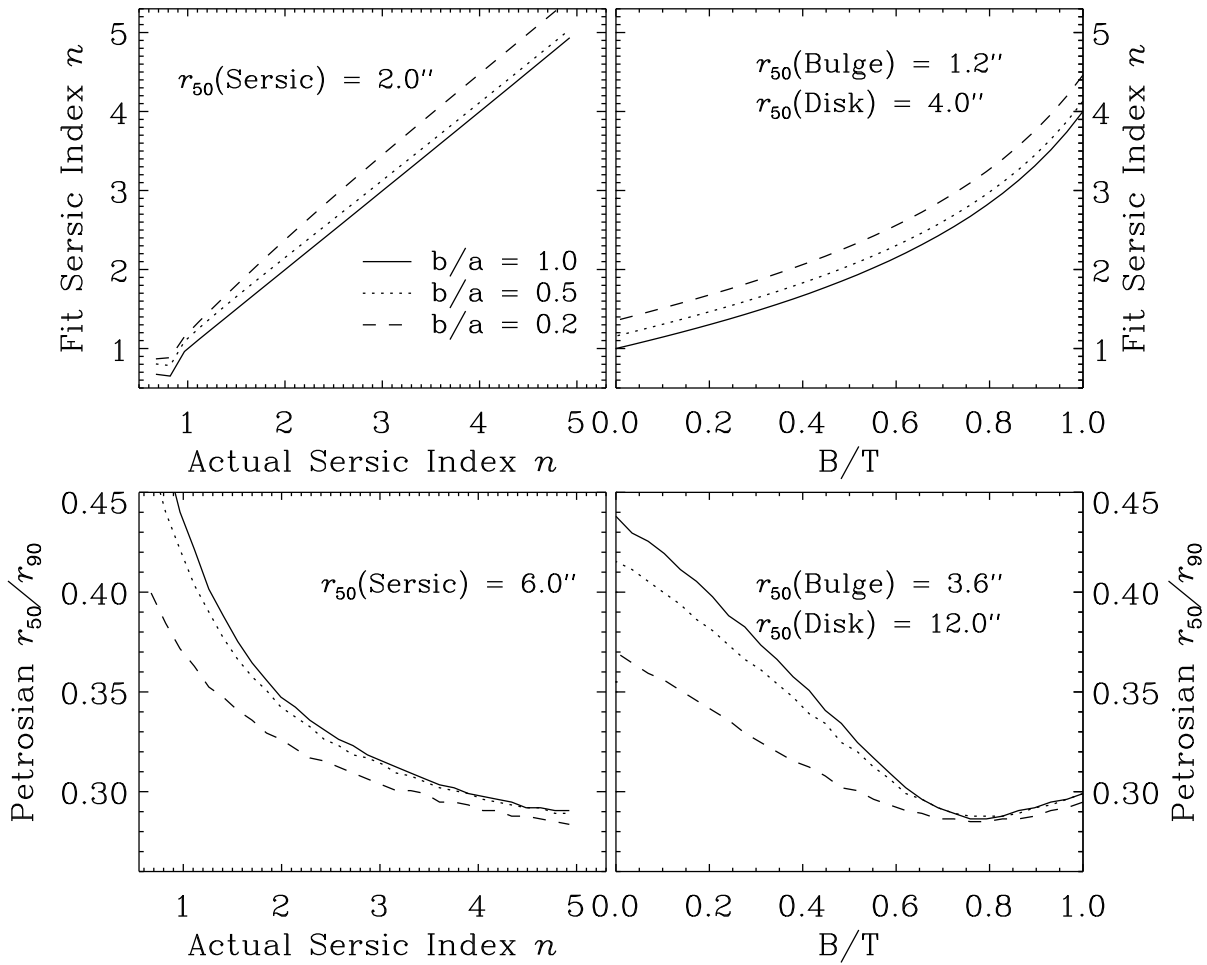}
\caption{\label{test_nfit_plot} Brief exploration of the meaning of
the \Sersic\  fits performed in this paper. We produced noiseless
simulations of galaxies, pixelized using 0.396$''$ pixels (the size of
the SDSS pixels), and applied gaussian seeing with $\sigma= 1.2''$. We
then extract radial profiles using the same annuli as used by the SDSS
{\tt photo} pipeline, and apply our \Sersic\  fit to the radial profile
(equally weighting all points in the profile). In the left column we
show the resulting index $n$ in the fit; in the right column we show
the Petrosian inverse concentration parameter $1/c= r_{50}/r_{90}$ for
the this best-fit \Sersic\  profile (just for comparison). We choose two
sorts of profiles: in the top panels, we show results for \Sersic\ 
profiles with various choices of \Sersic\  index $n$ but a fixed $r_{50}$
(as labeled); in the bottom panels, we show results for a bulge plus
disk galaxy, with $r_{50}$ for each component chosen as shown and a
varying ratio $B/T$ of total bulge flux to total disk flux. In both
cases we choose three different axis ratios, assuming
transparency. Deviations from axisymmetry under these conditions tends
to increase the concentration of the galaxy; on the other hand, one
might expect different behavior for opaque objects.}
\end{figure}

It is worth examining how our axisymmetric \Sersic\ fit is affected by
the non-axisymmetry of galaxies, and how it is related to other
measures of profile shape, such as the Petrosian concentration
parameter and the bulge-to-total ($B/T$) ratio. In the top panels of
Figure \ref{test_nfit_plot}, we investigate how our \Sersic\ fits
behave for galaxies with \Sersic\ profiles and galaxies which consist
of a de~Vaucouleurs bulge plus an exponential disk. In each case we
assume the galaxies are comparable in half-light radius to SDSS Main
Sample galaxies. We use three different axis ratios $b/a$ (as listed
in Figure \ref{test_nfit_plot}) by which we distort the galaxy image.
We then project that image onto a grid of pixels of size $0.396''$,
the size of SDSS pixels, apply single gaussian seeing with a standard
deviation of $1.2''$, and then extract radial profiles using a similar
scheme as used by SDSS {\tt photo}. Using the single gaussian model
for the seeing, we fit the \Sersic\ models to the resulting galaxy
profiles. We perfectly reconstruct the case of $b/a$ for the \Sersic\
profiles; for small axis ratios, we measure a higher concentration
than we would for a circular image with the same profile as along the
semi-major axis. For the bulge-plus-disk models, we find that the
\Sersic\ index $n$ is monotonically related to the $B/T$ ratio of the
bulge flux to the total flux.

The bottom panels of Figure \ref{test_nfit_plot} show the dependence
of the Petrosian inverse concentration parameter $r_{50}/r_{90}$ for
galaxies of the same shapes as, but larger radii than, the galaxies in
the top panels. This situation is meant to be similar to the
conditions under which galaxies were observed by \citet{shimasaku01a}
and \citet{strateva01a}, who observed large, bright galaxies less
affected by seeing than those in the SDSS Main sample. We find that
$r_{50}/r_{90}$ is closely related to \Sersic\ index and $B/T$.

We define the half-light surface brightness $r_{S,50}$ to be the
average surface brightness within the Petrosian half-light radius for
the \Sersic\ profile fit in the $\band{0.1}{i}$-band, in magnitudes in
one arcsec$^2$, $K$-corrected and corrected for cosmological surface
brightness dimming:
\begin{equation}
\mu_i \equiv m_{\mathrm{S},\band{0.1}{i}} + 2.5 \log_{10}(2 \pi
r_{\mathrm{S},50}^2) - 10 \log_{10} (1+z) - K_{\band{0.1}{i}}(z).
\end{equation} 
This measure of surface brightness is thus slightly different (in
general, higher) than the raw estimate calculated from the Petrosian
half-light radius output by the SDSS photometric pipeline, because it
is corrected for seeing.  Note that we have implicitly assumed that
the $K$-correction of the observation of surface brightness remains
unaffected by color gradients in the galaxy; because the
$K$-correction from the observed $i$ band to the $\band{0.1}{i}$ band
are expected to vary little with redshift and color ($< 0.2$
magnitudes) this approximation is not bad.

%

\subsection{Local Galaxy Density}
\label{density}

For each object in the sample we estimate the local galaxy density
$\rho$ as follows. First, we distribute a large number ($2 \times
10^6$) of Poisson random points within the SDSS volume, distributed in
redshift according to the selection function of the survey. The
selection function is determined by an evolving $\band{0.1}{r}$-band
Schechter function fit to the luminosity function of the following
form (following \citealt{lin99a}):
\begin{equation}
\Phi(L) dL = \phi_\ast \left(\frac{L}{L_\ast(z)}\right)^\alpha
\exp\left(- L/L_\ast(z)\right),
\end{equation}
where
\begin{equation}
L_\ast(z) = 10^{0.4 Q (z-0.1)} L_\ast(z=0.1).
\end{equation}
Expressed in magnitudes, this implies $M_\ast(z) = M_\ast(z=0.1) - Q
(z-0.1)$.  If $Q>0$, these formulae mean galaxies are brighter in the
past. We indeed find that in order to explain the dependence of number
counts on redshift in the survey, it is necessary to include the
possibility that galaxy luminosities evolve. We perform this fit in
the $\band{0.1}{r}$ band and find the parameters:
\begin{eqnarray}
\phi_\ast &=& 3.22 \times 10^{-2} \mathrm{~} h^3 \mathrm{~Mpc}^{-3},\cr
M_\ast &=& -20.50, \cr
\alpha &=& -1.01, \cr
Q &=& 1.97. 
\end{eqnarray}
We do not give error bars for this fit, nor do we recommend this fit
as our best estimate of the luminosity density or its evolution.  We
do not give error bars for this fit, nor do we recommend this fit as
our best estimate of the luminosity density or its evolution. A more
complete discussion of this topic is forthcoming in a separate
paper. However, this fit is good enough to use to estimate local
density.  Using this fit, the magnitude limits as a function of
position on the sky ($m_{r,\mathrm{min}}(\theta,\phi)$ and
$m_{r,\mathrm{max}}(\theta,\phi)$), and the spectroscopic sampling
fraction as a function of position on the sky ($f(\theta,\phi)$) we
can calculate the selection function at any point on the sky.

Given the galaxy catalog and the random catalog, we count the number
of galaxies $N_g$ (including the galaxy in question) and the number of
random points $N_r$ in a 8 $h^{-1}$ Mpc radius comoving sphere around
each galaxy. We define 
\begin{equation}
\rho \equiv \frac{N_g/\avg{N_g}}{N_r/\avg{N_r}},
\end{equation}
where $\avg{N}$ is the average expected number of objects in the
sphere. We use this as an estimate of the local density relative to
the mean. Because we use a spherical cell only 800 km s$^{-1}$ in
radius, this measure of density is affected somewhat by the velocity
dispersions in the deep potential wells of clusters, in the sense that
it will slightly underestimate the density in those regions. Also,
since it is centered on a galaxy, and galaxies are clustered on these
scales, its mean is greater than unity.

The local density is the only quantity considered here which is
measured with a large amounts of noise for each object. The average
``expected'' number of objects for a Poisson distribution in each cell
is about 14; the average actual number is about 37 (it is so much
larger because the variance of the overdensity on this scales is
around unity). This indicates that the average cell has errors of
$\sim 20\%$.  In particular, distant galaxies (which only have the
brightest galaxies at that redshift within 8 $h^{-1}$ Mpc) tend to
have a measurement of $\rho$ noisier than do close galaxies. One
cannot really consider the joint distribution of the local density
measured here and other galaxy properties, without accounting for the
noise in the measure of the local density (\citealt{blanton00a}).
Because of these effects, the galaxies which have the highest density
estimates are just the galaxies with the smallest expected number of
galaxies around them (and thus the highest expected noise in the
measurement). Furthermore, the galaxies with the smallest expected
number of galaxies around them are those at the highest redshift, and
are therefore the highest luminosity galaxies. This introduces a
correlation between the luminosity of a galaxy and the noise in our
estimate of the density. The galaxies with the highest density
estimates are high luminosity galaxies simply because of this
artificial correlation between luminosity and measurement noise. This
effect means that one cannot determine (from our measurements) the
density dependence of luminosity (or any property correlated with it
--- that is, all galaxy properties).

\begin{figure}
\figurenum{\fignum}
\plotone{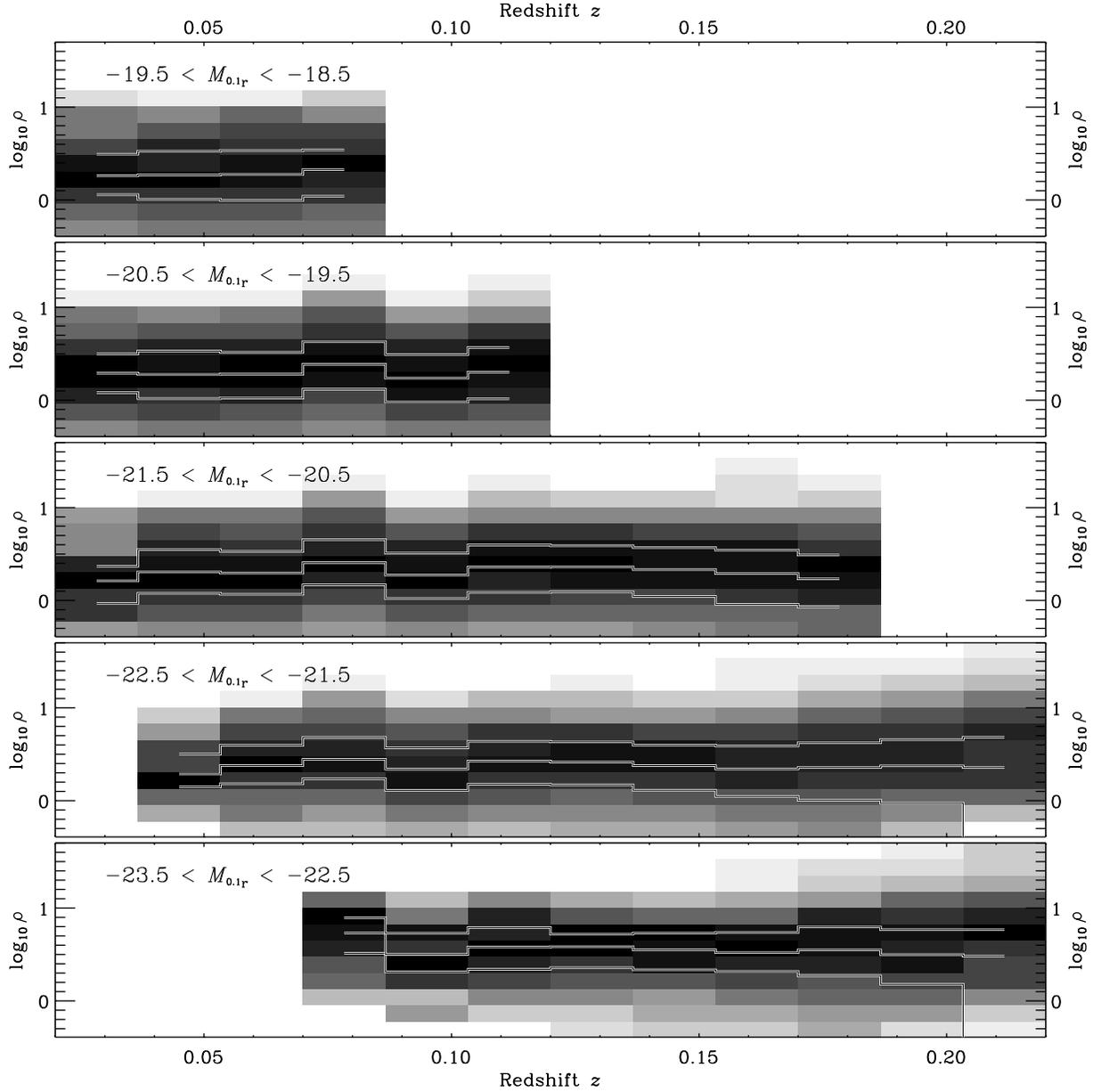}
\caption{\label{emg_density} Estimate of the local density within 8
$h^{-1}$ Mpc as a function of redshift, for a series of volume limited
samples with the absolute magnitude limits shown. As in Figure
\ref{emg_colors}, the greyscale is the conditional distribution of the
local density given the redshift, and the lines are the 10\%, 50\% and
90\% quantiles. The dependence of clustering on luminosity is evident.
The impact of Poisson noise on the local density estimates becomes
noticeable as the sample approaches $z\approx 0.2$, as the
distribution widens. However, the median local density appears nearly
constant, indicating that it is a reliable estimator of local
density.}
\end{figure}

Nevertheless, one can still meaningfully consider the median density
within bins of other galaxy properties, because the increased noise
for galaxies with higher mean distances does not affect the median
density. As proof of this assertion, Figure \ref{emg_density} shows
the density distribution as a function of redshift for several small
ranges of absolute magnitude. The lines are the 10\%, 50\%, and 90\%
quantiles of the distribution. Although at large distances the
distribution widens (the 10\% and 90\% quantiles decrease and
increase, respectively), the median remains remarkably constant,
indicating that it measures the density around galaxies consistently
at high redshift and at low redshift. This consistency means that even
when we divide galaxies into different groups which cover different
redshift ranges (such as dividing by luminosity) we can make a fair
comparison among the median local densities of the groups. This
consistency happens to be the justification for the common practice of
calculating the correlation function for galaxies of different
luminosities, in different redshift ranges.

%

\subsection{Contribution to Number Density}

To compute the global number and luminosity densities, it is necessary
to compute the number-density contribution $1/\Vmax$ for each galaxy,
where $\Vmax$ is the volume covered by the survey in which this galaxy
could have been observed, accounting for the flux, surface brightness,
and redshift limits as a function of angle (Schmidt 1968). This volume
is calculated as follows:
\begin{equation}
\Vmax = \frac{1}{4\pi}\,\int\mathrm{d}\Omega\,f(\theta,\phi)\, 
\int_{z_{\mmin}(\theta,\phi)}^{z_{\mmax}(\theta,\phi)}
\mathrm{d}z\,\frac{dV}{dz} \;,
\end{equation}
where $f(\theta,\phi)$ is described above and
$z_{\mmin}(\theta,\phi)$ and $z_{\mmax}(\theta,\phi)$
are defined by:
\begin{eqnarray}\displaystyle
z_{\mmin}(\theta,\phi) &=& \mmax(
z_{m,\mmin}(\theta,\phi), z_{\mu,\mmin}(\theta,\phi),
0.02)\cr
z_{\mmax}(\theta,\phi) &=& \mmin(
z_{m,\mmax}(\theta,\phi), z_{\mu,\mmax}(\theta,\phi),
0.22) \;.
\end{eqnarray}
The flux limits $m_{r,\mmin}(\theta,\phi)$ (equal to 14.5~mag across
the survey) and $m_{r,\mmax}(\theta,\phi)$ (approximately 17.77~mag)
implicitly set $z_{m,\minmax}(\theta,\phi)$ by:
\begin{eqnarray}\displaystyle
m_{r,\minmax}(\theta,\phi) &=& \fixMr +
\mathrm{DM}(z_{m,\minmax}(\theta,\phi))\cr
& & + K_\fixr(z_{m,\minmax}(\theta,\phi)) \;.
\end{eqnarray}
The surface brightness limits $\mu_{r,\mmin}(\theta, \phi)$ (in
practice, too high surface brightness to matter for this sample) and
$\mu_{r,\mmax}(\theta,\phi)$ (approximately 24.5~mag but variable
in a known way across the survey) implicitly set
$z_{\mu,\minmax}(\theta,\phi)$ by:
\begin{eqnarray}\displaystyle
\mu_{r,\minmax}(\theta,\phi) &=& \mu_{\band{0.1}{r}} +
10 \log_{10} (1+z_{\mu,\minmax}(\theta,\phi))\cr
& & + K_\fixr(z_{\mu,\minmax}(\theta,\phi)) \;.
\end{eqnarray}
Note that we have implicitly assumed that the $K$-correction of the
observation of surface brightness remains unaffected by color
gradients in the galaxy.  In practice, the surface brightness limits
only rarely affect the $\Vmax$ determination.

The function $f(\theta,\phi)$ is the SDSS sampling fraction of
galaxies as a function of position on the sky. The total sampling rate
of galaxies is computed separately for each region covered by a unique
set of spectroscopic survey tiles.  We adopt the nomenclature of the
2dF (\citealt{percival01a}): each such region is a ``sector'' (which
corresponds identically to the ``overlap regions'' defined in
\citealt{blanton01a}). That is, in the case of two overlapping tiles,
the sampling is calculated separately in three sectors: the sector
covered only by the first tile, the sector covered only by the second
tile, and the sector covered by both.  Each position on the sky is
thereby assigned a sampling rate $f(\theta,\phi)$ equal to the number
of redshifts of galaxy targets obtained in the region divided by the
number of galaxy targets in the region. In regions covered by a single
tile, typically $0.85<f<0.9$; in multiple plate regions, typically
$f>0.95$. These completenesses average to 0.92.

With $V_{\mathrm{max}}$, we calculate the global number and luminosity
density contributions of any group of galaxies as $\sum_i
1/V_{i,\mathrm{max}}$ and $\sum_i L_i/V_{i,\mathrm{max}}$.

\section{Results}
\label{results}

\subsection{Results for All Galaxies}

\subsubsection{Overview}

\begin{figure}
\figurenum{\fignum}
\plotone{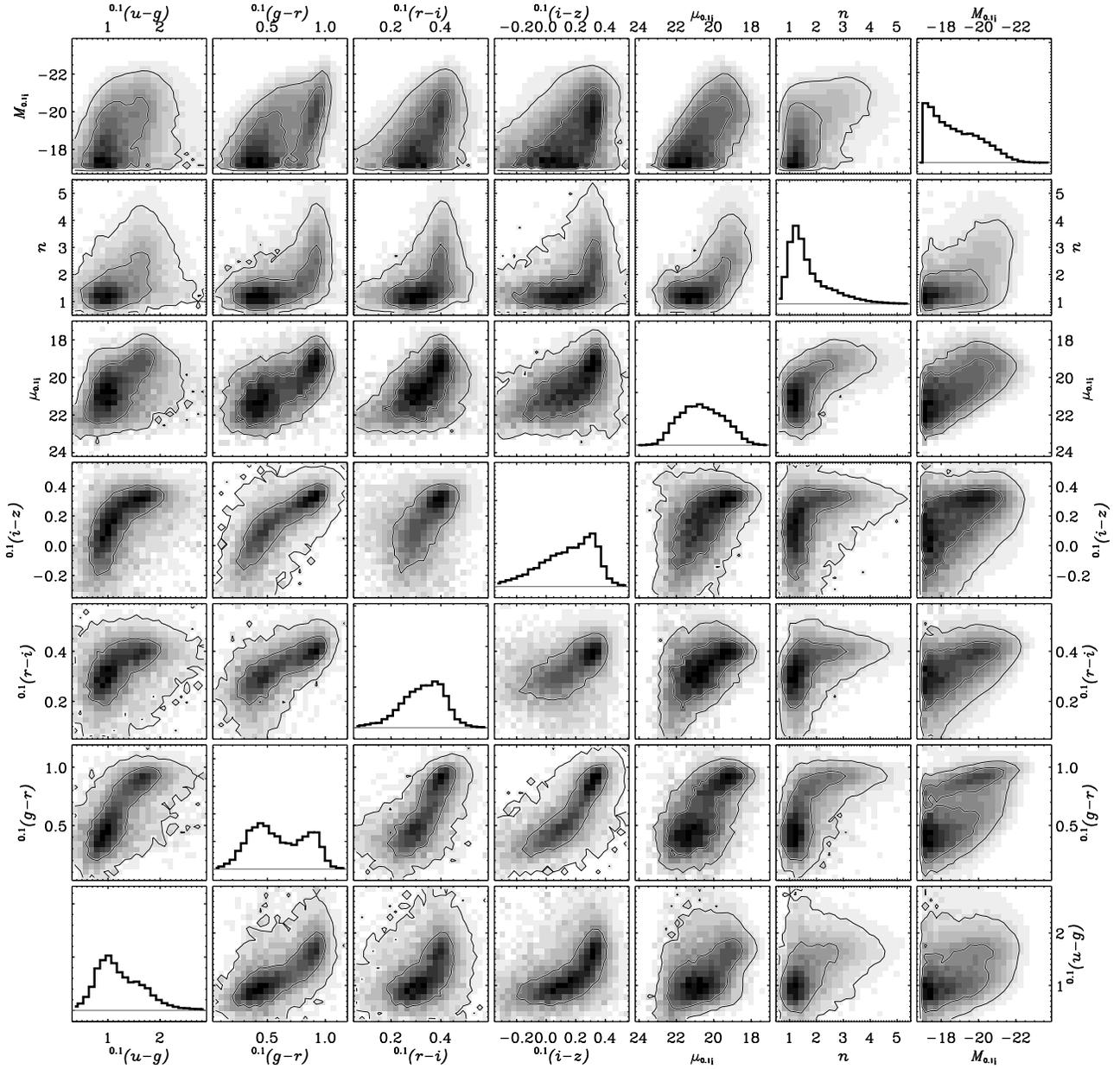}
\caption{\label{emg_num} Projection of the number density of
galaxies in our seven-dimensional space onto each pair of
dimensions. 
All images have a square-root stretch applied to increase the dynamic
range of the plot. Contours indicate the regions containing 68\% and
95\% of the total number density of galaxies in this sample.  The
upper and low triangles are identical mirror images.  The histograms
along the diagonal show the distribution of galaxies in each
dimension.}
\end{figure}

\begin{figure}
\figurenum{\fignum}
\plotone{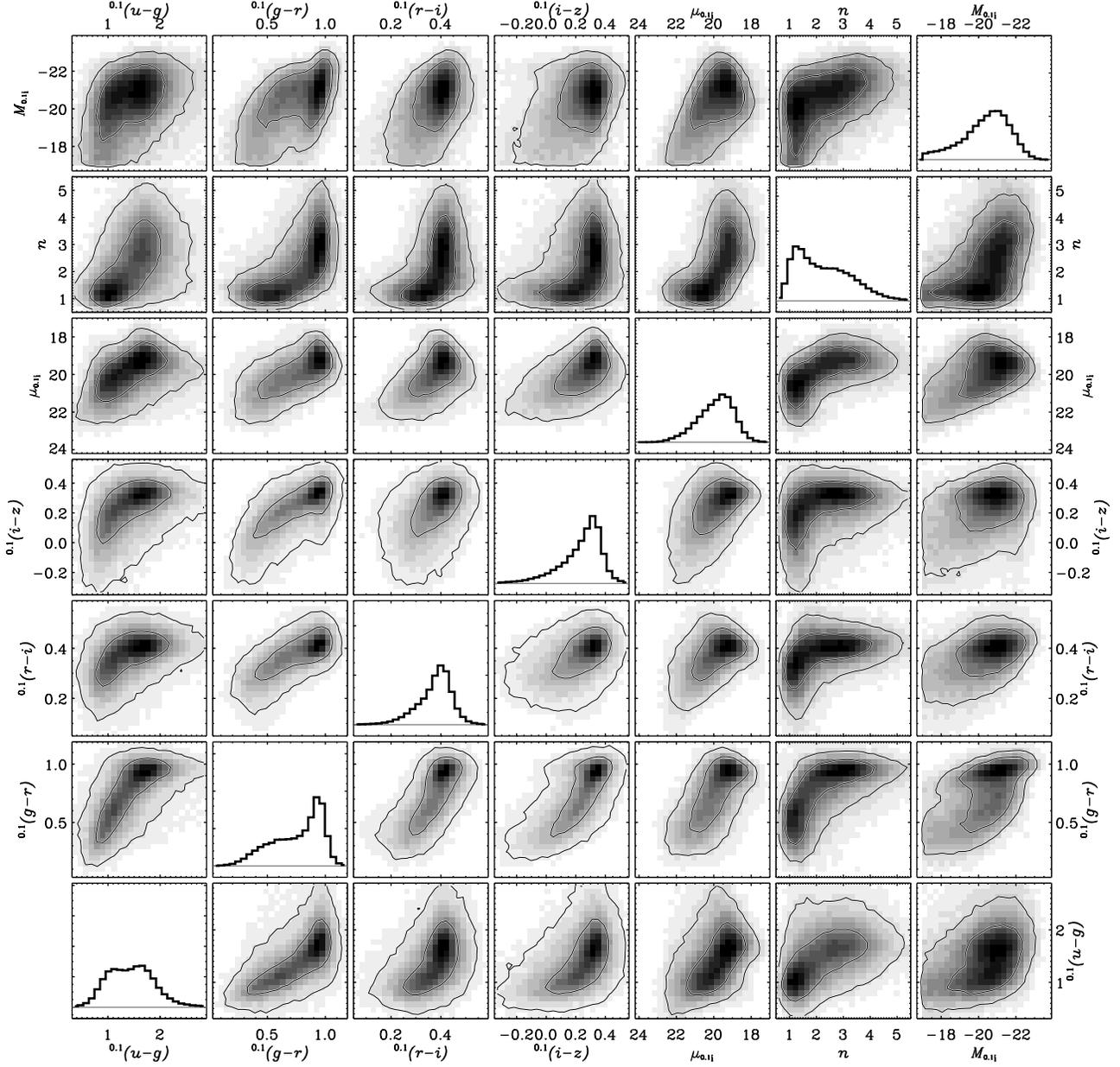}
\caption{\label{emg_lum} Same as Figure \ref{emg_num}, but
plotting luminosity density rather than number density. It is clear
from these plots that the luminosity density contribution from
galaxies below the absolute magnitude and surface brightness limits of
this sample is small.}
\end{figure}

\begin{figure}
\figurenum{\fignum}
\plotone{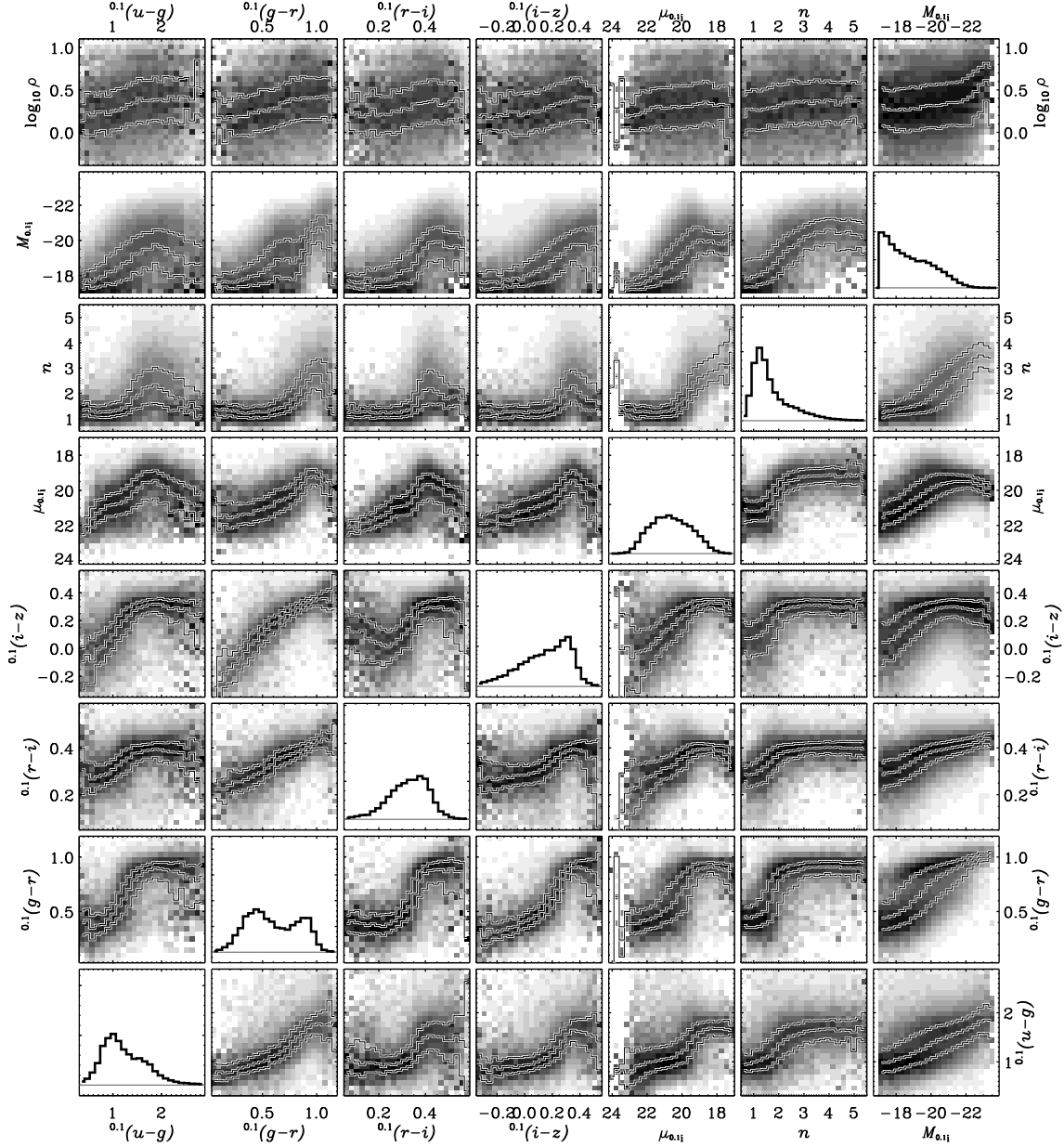}
\caption{\label{emg_num_cond} Similar to Figure \ref{emg_num}, but
showing the conditional distribution of each quantity on the vertical
axis with respect to the quantity on the horizontal axis.  Thus, the
upper left panel shows the conditional distribution $P(\log_{10}\rho |
M_{\band{0.1}{i}})$ while the lower right panel shows the conditional
distribution $P(\band{0.1}{(u-g)} | M_{\band{0.1}{i}})$. In other
words, these plots are equivalent to normalizing separately each
column of each panel in Figure \ref{emg_num}. The lines shown are
quartiles of the conditional number density distribution. While we
left out local density in Figures \ref{emg_num} and \ref{emg_lum}
because it is a noisy quantity and to calculate its distribution
requires accounting for this noise, we include it here since (as
concluded in the text) it is reasonable to look at the dependence of
the median local density on other galaxy properties.}
\end{figure}

Our primary results on the distributions of colors, surface
brightnesses, profile shapes, luminosities and local densities of
galaxies are shown in Figures \ref{emg_num}--\ref{emg_num_cond}. Each
of these plots is a multi-panel figure showing some form of the
bivariate distribution of every pair of galaxy properties in the
off-diagonal plots. The plots on the diagonal show the univariate
distribution of each quantity.

Figure \ref{emg_num} shows the projected number density distribution
for all pairs of seven of our parameters, excluding density (for
reasons described above). The \Vmax\ values calculated above are used
to calculate the contribution of each galaxy to the total number
density. The panels above and below the diagonal are mirror images of
each other.  The grey scale is proportional to the square-root of the
projected density in that plane. Contours in the plot represent loci
of constant projected density enclosing 68\% and 95\% of the density.
Figure \ref{emg_lum} shows the bivariate $\band{0.1}{i}$-band
luminosity density distribution between all of our parameters. This
figure is similar to Figure \ref{emg_num}, only now each object
contributes $L/\Vmax$ rather than $1/\Vmax$ to the density.

Figure \ref{emg_num_cond} shows the conditional number density
distribution of all eight of our parameters relative to seven of them
(only excluding density as an independent variable).  That is, the
greyscale represents the probability of a particular value of the
quantity on the vertical axis, given the value of the quantity on the
horizontal axis. This is equivalent to taking Figure \ref{emg_num},
and for each vertical column of pixels renormalizing by the total
density in that column.  This plot is useful when the dependence of
one quantity on another is strong in the regime where there are few
galaxies (such as at high density or high luminosity). The histograms
along the diagonal are the number density distributions of each
quantity, and are identical to the histograms in Figure \ref{emg_num}.

Some basic results are visible in these plots, which we review in the
following subsections. Many of these results have been noted in the
literature in the past. We will concentrate mostly on Figure
\ref{emg_num_cond}, in which many of the relationships are easiest to
see.

\subsubsection{Dependence on Luminosity}

The first simple result to point out from Figure \ref{emg_num_cond} is
the histogram in the right column and the second row from the top,
which shows the luminosity function in the \band{0.1}{i}-band. The
function is close to a Schechter function, although because the
vertical scale is linear, it may not appear familiar.  Now consider
the same panel in Figure \ref{emg_lum}, which shows the luminosity
density as a function of absolute magnitude. It is clear from this
plot that the luminosity density is almost entirely contained within
the range of absolute magnitudes shown ($-23.5 <M_{\band{0.1}{i}}
<-17.0$). We may further ask whether the surface brightness limits
(around $\mu_{\band{0.1}{i}} \sim 24.5$) greatly affect any estimate
of the luminosity density from this sample. The histogram in Figure
\ref{emg_lum} showing the luminosity density as a function of surface
brightness demonstrates that the luminosity density falls off long
before the surface brightness limit becomes important. An even clearer
demonstration of this fact is the panel showing the luminosity density
versus $M_{\band{0.1}{i}}$ and $\mu_{\band{0.1}{i}}$. This panel shows
that even at the lowest luminosities in the sample, the surface
brightness limit excludes very little luminosity density. These
results are in agreement with \citet{blanton01a} (using early SDSS
data) and \citet{cross01a} (using early 2dFGRS data). A more careful
analysis of the luminosity function and its evolution from this sample
is being carried out by Blanton et al. (in preparation).

\begin{figure}
\figurenum{\fignum}
\plotone{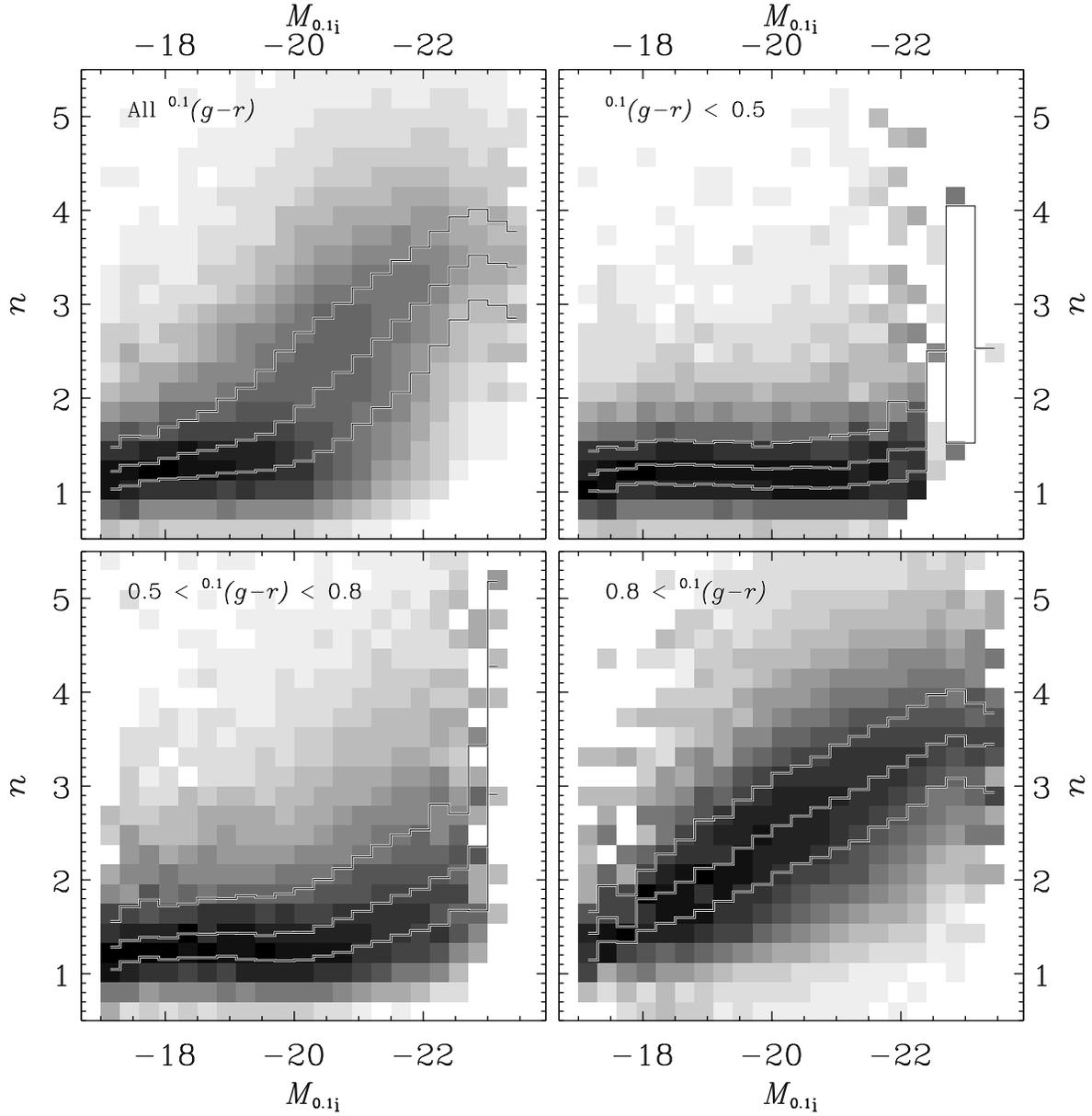}
\caption{\label{emg_nMgmr} Similar to Figure \ref{emg_num_cond}, 
but restricting only to the dependence of \Sersic\ index $n$ on
absolute magnitude, for several ranges of $\band{0.1}{(g-r)}$ color,
as labeled. Blue galaxies are overwhelmingly exponential; red galaxies
become more and more concentrated as they become more luminous.}
\end{figure}

Next, let us follow the right column of Figure \ref{emg_num_cond} down
from the top. Each panel below the top panel shows the conditional
dependence of a galaxy parameter on luminosity. These panels show many
of the familiar relationships of galaxy properties to luminosity from
previous observations. In sequence from top to bottom:
\begin{enumerate}
\item Local density is a strong function of luminosity. The most
luminous galaxies exist preferentially in the densest regions of the
universe. Density increases with luminosity for all absolute
magnitudes above $M_{\band{0.1}{i}} = -18$. At lower luminosities, the
local density starts to increase again, which is potentially related
to the existence of dwarf spheroidal galaxies in clusters. This result
agrees with results from CfA (\citealt{hamilton88a}), the Optical
Redshift Survey (hermit96a), as well as the recent results of
\citet{zehavi02a} in the SDSS and \citet{norberg02a} in the 2dFGRS on
the dependence of the correlation function amplitude on galaxy
luminosity.
\item Highly luminous galaxies are more concentrated, and thus have
higher \Sersic\ indices, than lower luminosity galaxies. This trend is
consistent with high luminosity galaxies being preferentially
``early-type.'' However, even for the highest luminosity galaxies, the
de~Vaucouleurs profile ($n\sim 4$) is always more concentrated than
more than 75\% of the galaxies; the total number density of galaxies
with $n>3$ in our absolute magnitude range is $\sim 5\%$. Thus, either
the number density fraction of ``true ellipticals'' is small in this
absolute magnitude range or many ``true ellipticals'' are not de
Vaucouleurs profile galaxies.  That the \Sersic\ profile in most cases
provides a better fit to the surface brightness profiles of
morphologically-classified elliptical galaxies than the de~Vaucouleurs
profile has been shown previously by several groups
(\citealt{prugniel97a,caon93a}). We can try to compare to these
results by dividing our sample by color. Figure \ref{emg_nMgmr} shows
the dependence of $n$ on absolute magnitude for three ranges of
$\band{0.1}{(g-r)}$. The \Sersic\ index $n$ is clearly independent of
luminosity for very blue galaxies (which are close to
exponential). For very red galaxies the \Sersic\ index is a strong
function of luminosity (paralleling the dependence of \Sersic\ index
on luminosity found by \citealt{prugniel97a} and \citealt{caon93a} for
morphologically elliptical galaxies).
\item Luminosity and surface brightness are positively correlated at
low luminosities, while at extremely high luminosity surface
brightness declines with luminosity. As we will see below, this occurs
because the surface brightnesses of highly concentrated galaxies,
which tend to be highly luminous, decline with luminosity, while the
surface brightnesses of exponential galaxies, which tend to be low
luminosity, increase with luminosity. Note that constant size as a
function of surface brightness would be a 45 degree line on the
luminosity-surface brightness plot. At all luminosities, the surface
brightness increases more slowly than luminosity, indicating that the
median size of galaxies increases strongly with luminosity, especially
for the most luminous galaxies.
\item The colors of galaxies depend strongly on luminosity, in the
sense that more luminous galaxies are redder. The color-magnitude
relation is most obvious in \band{0.1}{(g-r)}, where separate red and
blue populations are evident. This bimodality has also been noted in
the observed $u-r$ colors by \citet{strateva01a} and in the D4000
spectroscopic measurements of \citet{kauffmann02a}. 
\end{enumerate}
In short, there are strong dependencies of all parameters on
luminosity. 

\subsubsection{Dependence on Profile Shape}

The number density distribution of $n$, the \Sersic\ index, shows a
strong peak at $n=1$ in Figures \ref{emg_num} and \ref{emg_num_cond},
indicating that exponential galaxies are extremely common in the
universe. Note that there is no peak in the number density at $n=4$,
which corresponds to a de~Vaucouleurs profile, indicating that de
Vaucouleurs galaxies lie along a continuum of galaxy concentration.
Naturally, this result is subject to the limitations of our fit ---
which is based on an azimuthally averaged profile and uses a double
gaussian approximation to the seeing. On the other hand, Figure
\ref{test_nfit_plot} suggests that the effects of inclination are more
likely to increase the observed \Sersic\ indices. To test the effects
of inclination more empirically, we can restrict our sample to
galaxies which are nearly face on (axis ratio greater than 0.8),
according to the {\tt photo} pipeline's {\tt ab\_exp} and {\tt
ab\_dev} measurements of the best fit axis ratios assuming an
exponential or de~Vaucouleurs model, respectively. Using either {\tt
ab\_exp} or {\tt ab\_dev} as a measure of axis ratio does not lead the
\Sersic\ fits to be closer to $n=4$ or for any separate
``concentrated'' peak to appear in the distribution of \Sersic\ index
$n$. Similarly, if we restrict our sample to galaxies which have
$\chi^2$ less than $N_d + \sqrt{N_d}$ (where $N_d$ is the number of
degrees of freedom in the fit), we also don't find any peak at a de
Vaucouleurs profile. The only subset of galaxies which we have found
in our parameter space whose \Sersic\ index even close to
``typically'' $n=4$ are very red ($\band{0.1}{(g-r)} > 0.8$) and very
luminous ($M_{\band{0.1}{i}} < -22.5$), which one can see in Figure
\ref{emg_nMgmr}; the actual number of such galaxies is very
tiny. Thus, if there is a single parameter one can measure about
galaxies which indicates whether it would be classified
morphologically as an elliptical, and the typical such elliptical
really has $n=4$, that parameter is not any of those considered here.

The second column from the right in Figure \ref{emg_num_cond} shows
the dependence of each quantity on $n$, the \Sersic\ profile index. From
top to bottom:
\begin{enumerate}
\item Local density does not depend on \Sersic\ index as much as it
depends on luminosity, although exponential galaxies are in somewhat
lower density regions than are concentrated galaxies (about 25\%
less overdense).
\item Galaxies with $n\approx 4$ have a luminosity function peaked at
about $M_{\band{0.1}{i}} \approx -20.5$. The peak luminosity declines
as the profile shape becomes closer and closer to exponential ($n=1$),
to the extent that it is clear that the peak of the luminosity
function for pure exponential galaxies is well below our absolute
magnitude limit of $-17$. Due to these trends, the luminosity density
distribution of $n$ in Figure \ref{emg_lum} is far flatter than the
number density distribution, such that concentrated galaxies
contribute a much larger fraction of the luminosity density than they
do of the number density (for a more detailed analysis of this
phenomenon, see \citealt{hogg02a}).
\item The half-light surface brightnesses of exponential galaxies
($n\sim 1$) are distributed around $\mu_{\band{0.1}{i}} \sim 21$ while
the surface brightnesses of concentrated galaxies are centered at
much higher surface brightness ($\mu_{\band{0.1}{i}} \sim 19$).
\item The galaxies which are nearly exponential tend to be very blue
in all colors
($\band{0.1}{(g-r)}\sim 0.4$). For $n>3$, galaxies are extremely
homogeneous: constant and red colors, constant and high surface
brightness, and constant average density.
\end{enumerate}

\subsubsection{Dependence on Surface Brightness}

The third column from the right in Figure \ref{emg_num_cond} shows the
dependence of each quantity on $\mu_{\band{0.1}{i}}$;
\begin{enumerate}
\item Density is only a weak function of surface brightness on scales
of 8 $h^{-1}$ Mpc. Note that \citet{zehavi02a} investigate the
dependence of the correlation function on surface brightness; they
find little dependence on these scales, but a strong dependence on
smaller scales.
\item Paralleling the results for the \Sersic\ index, the luminosity
function of high surface brightness objects is peaked at high
luminosity, while at low surface brightness the luminosity function
continues to increase until it passes our low luminosity limit. This
result is similar to that found in SDSS commissioning data by
\citet{blanton01a} and in the 2dFGRS by \citet{cross01a}. 
\item The profiles of high surface brightness galaxies are
concentrated, while the profiles of most galaxies at lower surface
brightness than $\mu_{\band{0.1}{i}} \sim 20$ are consistent with
exponential. 
\item The highest surface brightness objects are uniformly red, while
fainter than $\mu_{\band{0.1}{i}} \sim 20$, surface brightness and
color are positively correlated --- the lower surface brightness a
galaxy is, the bluer it tends to be. 
\end{enumerate}

\subsubsection{Dependence on Color}

The four left columns in Figures \ref{emg_num}--\ref{emg_num_cond}
show the relation between the four optical colors of galaxies and the
other parameters. In Figure \ref{emg_num}, two distinct peaks in the
number density distribution are visible, most notably in the relation
between $\band{0.1}{(u-g)}$ and $\band{0.1}{(g-r)}$. This result
corresponds closely to the work of \citet{strateva01a}, which found a
double peaked distribution of (observer-frame) $u-r$ color. The four
left columns in Figure \ref{emg_num_cond} show the dependence of all
properties on the colors:
\begin{enumerate}
\item Density is a strong function of all colors. One expects this
trend, based on the separate correlations of galaxy morphological
classification with color (\citealt{roberts94a} and references
therein) and environment (for example, \citealt{dressler80a}).
\item The luminosity function of galaxies which have
$\band{0.1}{(g-r)} \sim 1.0$ and $\band{0.1}{(r-i)} \sim 0.45$ is
peaked at high luminosity ($M_{\band{0.1}{i}} \sim -20.5$). The
luminosity function of bluer galaxies increases right up to our lower
luminosity limit. Redder than $\band{0.1}{(g-r)} \sim 1.0$ or
$\band{0.1}{(r-i)} \sim 0.45$, the peak luminosity declines as
well. This trend indicates that the reddest galaxies are a different
population than those right around $\band{0.1}{(g-r)} \sim 1.0$.
\item Blue galaxies have preferentially exponential profiles. The
concentration of galaxy profiles is highest for galaxies with
$\band{0.1}{(g-r)} \sim 1.0$ or $\band{0.1}{(r-i)} \sim 0.45$
(although, again, the median never reaches $n\sim 4$) and declines at
redder colors again. Again, this trend suggests that the reddest
galaxies are a special population; in particular, the fact that they
tend to have profiles closer to exponential suggests that they are
edge-on spiral galaxies reddened by dust. This speculation is
supported by the fact that the axial ratios measured by {\tt photo}
for these objects are low ($\sim 0.2$--$0.3$).
\item The half-light surface brightness increases for redder galaxies,
again until $\band{0.1}{(g-r)} \sim 1.0$ or $\band{0.1}{(r-i)} \sim
0.45$, at which point the surface brightness declines again. This
trend may also be partly due to the fact that the reddest galaxies are
probably dusty spirals; however, it is also due to the trend that at
the higher luminosities, where galaxy colors are redder, galaxy
half-light surface brightnesses tend to get lower with luminosity.
\end{enumerate}
Furthermore, the colors are clearly highly correlated among
themselves. In particular, measuring $\band{0.1}{(g-r)}$ constrains
the other colors to very small ranges.

\subsection{Dividing Galaxies by \Sersic\ Index}
\label{sersicdivide}

\begin{figure}
\figurenum{\fignum}
\plotone{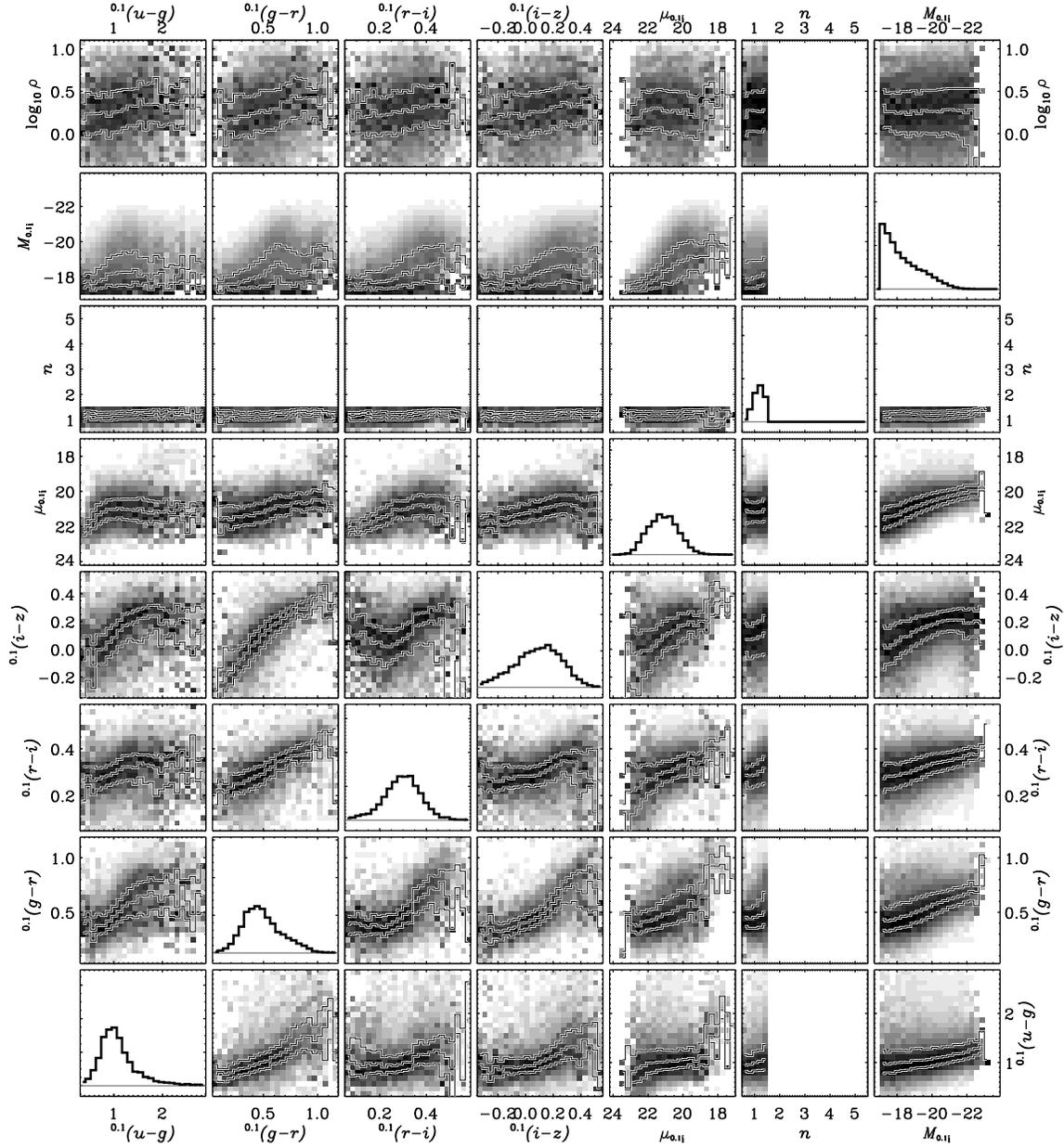}
\caption{\label{emg_n0_num_cond} Similar to Figure
\ref{emg_num_cond}, but only including galaxies with $n<1.5$ (the
exponential galaxies).}
\end{figure}

We have so far explored only two-dimensional projections of an
eight-dimensional space of parameters. It is interesting to explore a
third dimension of this space. To do so, we divide the galaxies into
two groups according to their \Sersic\ index.  We choose an
exponential group ($n<1.5$) and a concentrated group ($n> 3$). (The
results for the intermediate group look like a combination of the
contributions from each group separately).  Figures
\ref{emg_n0_num_cond} and \ref{emg_n2_num_cond} show the conditional
distribution of all our galaxy properties on all of the others for the
exponential and concentrated groups, respectively. It is easy to
identify which group of galaxies is considered from the panels
involving the \Sersic\ index --- for example, there is clearly no data
for $n>1.5$ in Figure \ref{emg_n0_num_cond}.

The diagonal panels in each plot (which can be inferred from Figure
\ref{emg_num}) demonstrate that the \Sersic\ index does indeed divide
galaxies into red, high surface brightness, luminous galaxies and
blue, low surface brightness, underluminous galaxies. In particular,
the luminosity function of the exponential group clearly is rising
steeply through our low luminosity limit, while the luminosity
function of the concentrated group is peaked at high luminosity.

Constrasting the conditional distributions of the exponential group in
Figure \ref{emg_n0_num_cond} to those of the concentrated group in
Figure \ref{emg_n2_num_cond} reveals that the properties of these
populations have very different interrelationships. Most strikingly,
the color-magnitude diagrams of these two classes separate
extraordinarily well, into a tight red sequence (probably
corresponding closely to the color--magnitude relation for
morphologically elliptical galaxies \citealt{baum59}) and a less
tight, but undeniable, blue sequence.

Similarly, the dependence of surface brightness on luminosity differs
markedly for the two classes. The surface brightness of exponential
galaxies increases with luminosity (as does their size), while the
surface brightness of concentrated galaxies decreases with their
luminosity (while their size increases even more strongly). Another
way of looking at the same issue is to consider the dependence of the
physical half-light radius on absolute magnitude; Figure
\ref{emg_r50Mn} shows this relationship for three ranges of \Sersic\
index $n$. Again, this figure shows that galaxy size closely
correlates with luminosity, with a stronger dependence for de
Vaucouleurs than for exponential galaxies. These results are
comparable to those in Figure 44 of \citet{bernardi02q}; if one simply
inverts the slope of the relationship between $r_{50}$ and $L$ (which
yields a correct result if the relationship between the two is
sufficiently linear and deterministic) we find $L_i\propto
r_{i,50}^{-1.67}$, close to their result. These results also hold for
the dependence of surface brightness on color; surface brightness
correlates with color for exponential galaxies, but peaks at around
$\band{0.1}{(g-r)}\sim 1.0$ for concentrated galaxies. The results for
concentrated galaxies are in good agreement with the work of
\citet{binggeli84a} (based on morphologically elliptical galaxies).

Some of the most interesting results from these figures are the
dependence of local density on galaxy properties for each type of
galaxy profile taken separately.  For example, the local density of
exponential galaxies does not depend on surface brightness; for de
Vaucouleurs, local density {\it decreases} slightly with surface
brightness, a fact probably related to the decrease of luminosity with
surface brightness for concentrated galaxies.  It is only that higher
surface brightness galaxies are more concentrated, and that
concentrated galaxies are on average in higher density regions, that
keeps the dependence of density on surface brightness flat for all
galaxies.

The dependence of the local density on luminosity has a similar {\it
shape} for concentrated galaxies and exponential galaxies, in the
range of luminosities in which there is data for both types (less
luminous than about $M_{\band{0.1}{i}} \sim -22$). However,
concentrated galaxies are naturally in slightly denser regions at
any given luminosity.  Meanwhile, for both concentrated and
exponential galaxies, density depends strongly on color.

\section{Fitting Median Properties as a Function of Luminosity}

\begin{figure}
\figurenum{\fignum}
\plotone{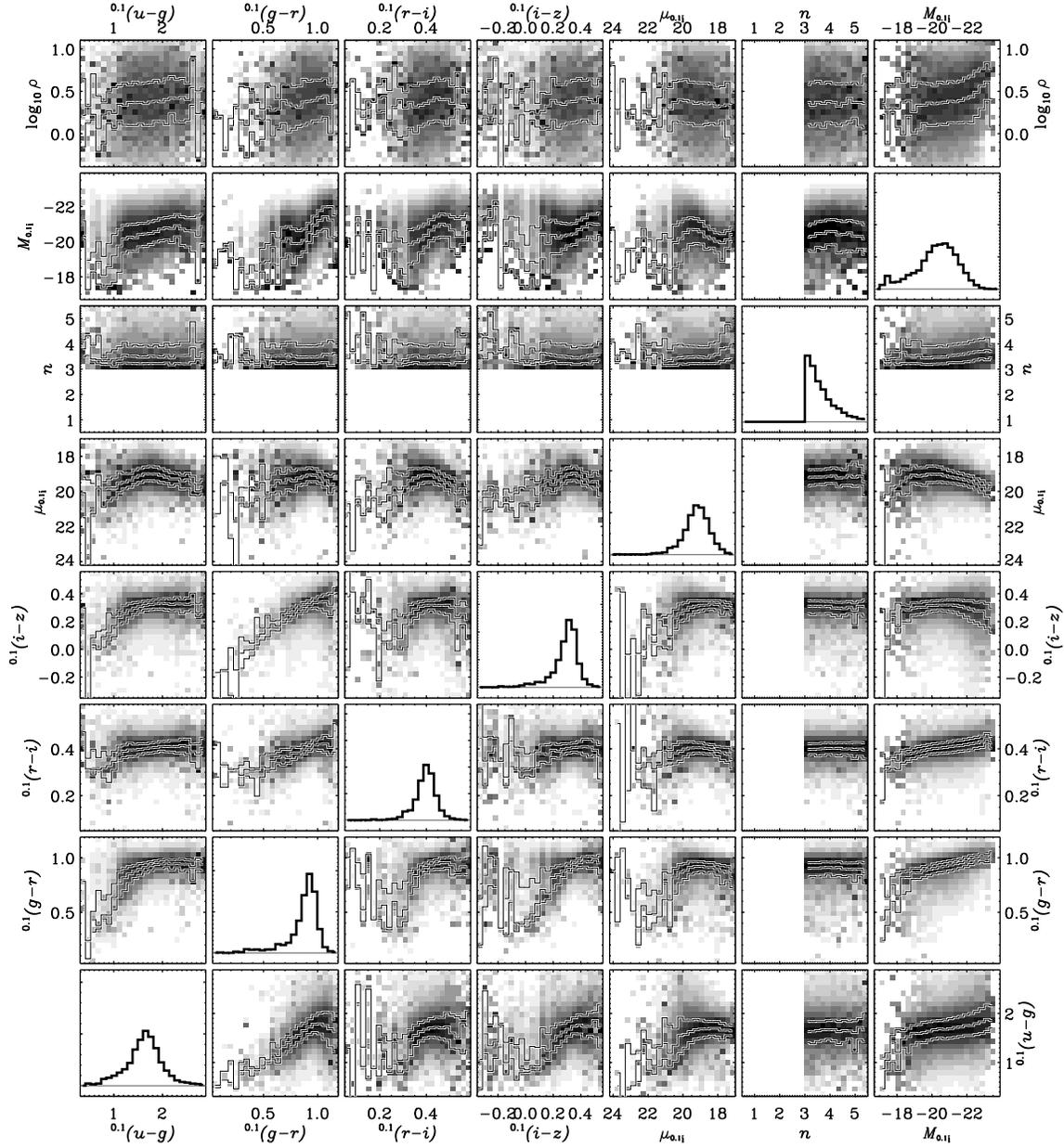}
\caption{\label{emg_n2_num_cond} Similar to Figure
\ref{emg_num_cond}, but only including galaxies with $n>3.0$ (the
concentrated galaxies).}
\end{figure}

\begin{figure}
\figurenum{\fignum}
\plotone{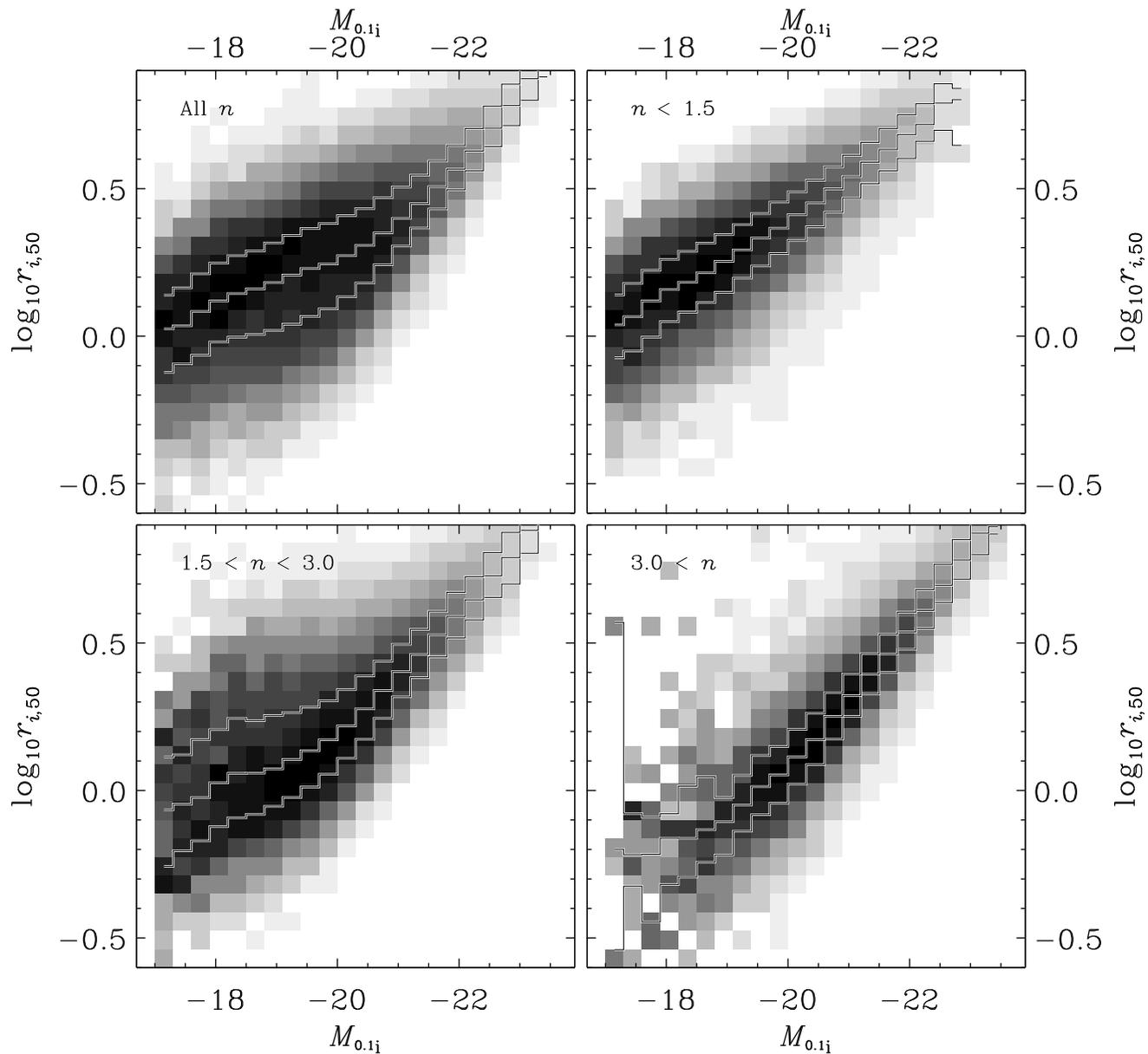}
\caption{\label{emg_r50Mn} Similar to Figure \ref{emg_nMgmr}, only
showing the distribution of the \Sersic\ half-light radius $r_{50}$ (in
units of $h^{-1}$ kpc) as a function of absolute magnitude for several
ranges of \Sersic\ index. The results here are simply another
expression of the relationship between absolute magnitude and
half-light surface brightness shown in Figures \ref{emg_num_cond},
\ref{emg_n0_num_cond}, and \ref{emg_n2_num_cond}.}
\end{figure}

Figures \ref{emg_n0_num_cond} and \ref{emg_n2_num_cond} show that
exponential and concentrated galaxies have colors, surface
brightnesses, and local densities which are distinct from one another
and which depend on galaxy luminosity.  To quantify these relations,
Figure \ref{cmrplot} shows the $1/\Vmax$-weighted median properties of
both types as the boxes for exponential (thin lines) and
concentrated (thick lines) galaxies (these lines are just the same
as the median lines in Figures \ref{emg_n0_num_cond} and
\ref{emg_n2_num_cond}). The statistical uncertainties in the median
containing 68\% of the probability are shown as vertical lines (the
width of the distribution is of course much larger).

We have fit these median relations in the following manner.  First, we
add a minimum variance in quadrature to the uncertainties on the
points, which is $(0.02)^2$ for the colors and surface brightness, and
is $(0.1)^2$ for density. We do so in order that our $\chi^2$
minimization does not obsessively fit the few points with extremely
tiny statistical uncertainties (who could easily be off by a small
amount due to systematic errors in the data and our processing). Next,
we fit a cubic polynomial to the resulting set of points and
uncertainties. To be explicit, for property $p$ we fit the function
\begin{equation}
\label{median_fit}
p = p_0 
+ p_1 (M_{\band{0.1}{i}} + 21)
+ p_2 (M_{\band{0.1}{i}} + 21)^2
+ p_3 (M_{\band{0.1}{i}} + 21)^3,
\end{equation}
where $M_{\band{0.1}{i}}$ is expressed in absolute magnitudes, the
surface brightness is in magnitudes in 1 arcsec$^2$, the colors are in
magnitudes, and $\log_{10} \rho$ is unitless.  We center the fit at
$M_{\band{0.1}{i}} = -21$ because that is near the center of our
luminosity range.  The resulting curves are shown in Figure
\ref{cmrplot} as continuous lines. They are all reasonable fits to the
data over the range we have fit. The parameters $p_i$ are listed in
Table \ref{cmrtable} for each property and galaxy profile type. We
also list the standard deviations $\sigma$ defined by the distribution
of galaxies around the best fit curves, both in Figure \ref{cmrplot}
and in Table \ref{cmrtable}. However, as can be seen in Figures
\ref{emg_n0_num_cond} and \ref{emg_n2_num_cond}, gaussians of constant
width are by no means adequate descriptions of the distributions
around the median.

These relations may be useful for comparison to theory or to other
observations. They allow one to predict the median properties of an
exponential or concentrated galaxy given only the absolute
magnitude. We caution that because there is significant scatter in all
of these relations, one cannot reverse {\it these} fits to predict
absolute magnitude from other galaxy properties.

\section{Accessing the Two-dimensional Distribution Data}

The two-dimensional projections of the number and luminosity density
data used to create Figures \ref{emg_num}, \ref{emg_lum},
\ref{emg_nMgmr}, \ref{emg_n0_num_cond}, \ref{emg_n2_num_cond}, and
\ref{emg_r50Mn} are available in FITS format from the electronic
version of this article {\bf [NOTE TO EDITOR: These electronic tables
will be required; for now these can be found at \\
{\tt http://wassup.physics.nyu.edu/manyd/paper/}]}.

The FITS file each have a single HDU containing three columns and one
row. The first column contains $M$ names (in string format) of the
quantities listed in that file (corresponding to the $x$ and $y$ axes
of the figures). The second column contains the ranges over those $M$
quantities included in the figure (the first element of each range
indicates the position of the left edge of the left-most pixel, and
the second element of each range indicates the position of the right
edge of the right-most pixel). The third column is a four dimensional
array of dimensions $(N,N,M,M)$ where $M$ is the number of quantities,
and $N$ is the number of pixels on a side of each image. Each $N\times
N$ subarray contains the data values used to create the greyscales in
the corresponding figure of this paper.  The values in the subarray
are per unit ``quantity $m_1$'' per unit ``quantity $m_2$'', where
$m_1$ and $m_2$ represent the position in the $(M,M)$ sized dimensions
in the four dimensional array. 

\section{Conclusions and Future Work}
\label{conclusion}

In this paper, we have presented the number and luminosity density
distributions of photometric galaxy properties as measured by the
SDSS. Galaxy properties are highly correlated. Here are our main
conclusions:
\begin{enumerate}
\item
The most luminous galaxies comprise a homogeneous red, highly
concentrated, high surface brightness population, and reside in
locally dense regions. Underluminous galaxies are less homogeneous,
but in general are bluer, less concentrated, lower surface brightness,
and reside in less dense regions. These results are in qualitative
agreement with previous work.
\item 
The relations among galaxy luminosities, surface brightnesses, and
colors separate neatly once one has separated them into concentrated
and unconcentrated (exponential) groups. Some of these relationships
are well-known (such as the color--magnitude relation of concentrated
galaxies; \citealt{baum59}) while others are less commonly discussed
(such as the color--magnitude relation for exponential galaxies).
\item Local density is a strong function of luminosity, at least for
the most luminous galaxies, and on color for galaxies of all
colors. This dependence of clustering on galaxy type has been
quantified many times previously, early on in the work of
\citet{oemler74a}, later in the cluster studies of
\citet{dressler80a}, and also on larger scales by
\citet{davis76a,giovanelli86a,santiago92a,guzzo97a,blanton00a}, and
many others.
\end{enumerate}

In addition, we make several more minor observations about the
distribution of galaxy properties:
\begin{enumerate}
\item 
Optical galaxy colors all correlate very strongly with
\band{0.1}{(g-r)} color. The distribution of \band{0.1}{(g-r)} is
double-peaked. 
\item 
Most of the luminosity density in the universe is contained in
galaxies within our range of galaxy luminosities ($-23.5 <
M_{\band{0.1}{i}} < -17.0$).
\item 
The very reddest galaxies (which are small in number) are in optical
colors exponential galaxies, not concentrated galaxies.
\item
For all types of galaxies, size increases with luminosity. For
concentrated galaxies, surface brightness decreases with luminosity
for the most luminous galaxies. For exponential galaxies, surface
brightness increases with luminosity.
\end{enumerate}

\begin{figure}
\figurenum{\fignum}
\epsscale{0.85}
\plotone{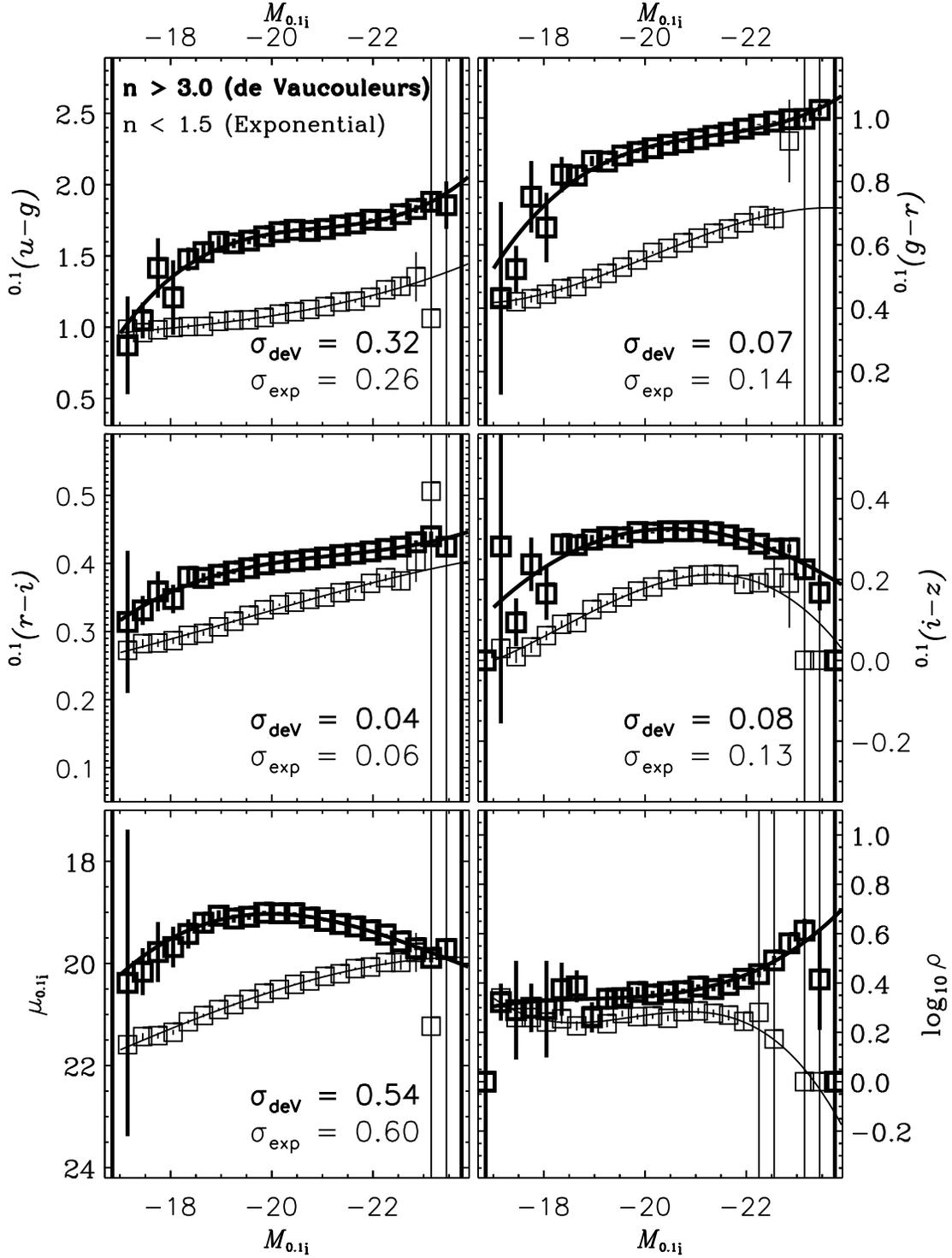}
\epsscale{1.0}
\caption{\label{cmrplot} Median galaxy properties for concentrated
galaxies ($n>3$; thick lines) and exponential galaxies ($n<1.5$; thin
lines). The measured data points are the boxes, with uncertainties
shown as vertical bars. The cubic polynomial fits of the form of
Equation \ref{median_fit} are the continuous lines; the parameters of
these fits are listed in Table \ref{cmrtable}. The $1\sigma$ deviation
around these fits are listed in each panel (and listed in Table
\ref{cmrtable}). The separation between exponential and concentrated
galaxies is clear.}
\end{figure}

Simple quantitative expressions of some of these results are contained
in the parametric fits to the conditional distributions of Figure
\ref{cmrplot} and Table \ref{cmrtable} as well as the two-dimensional
projections given in the associated electronic tables. However, in
future work we may characterize the distribution using a full
seven-dimensional model of the distribution of properties.

A potentially important effect to account for in future analyses is
the passive evolution of galaxy luminosities. Measurements of this
evolution in the SDSS suggest that galaxy luminosities are brighter in
the past (a trend of $4.2$, $2.0$, $1.6$, $1.6$, and $0.8$ magnitudes
per unit redshift in the \band{0.1}{u}, \band{0.1}{g}, \band{0.1}{r},
\band{0.1}{i}, and \band{0.1}{z} bands). While the difference in
absolute magnitude over the whole redshift range is quite small (0.3
magnitudes) compared to the dynamic range in $\band{0.1}{i}$-band
absolute magnitude, some of the colors may be affected
significantly. Because the survey is flux limited, this effect can, in
principle, alter the slope of our measurement of the color-magnitude
relationship. For the trends quoted here, accounting for evolution
would make the measured slope steeper, because the high luminosity
galaxies are observed primarily at high redshift, and are observed to
be bluer than they would be at the lower redshifts of the lower
luminosity galaxies to which we are comparing them. For example,
\citet{bernardi02q} have found such evolution of the $g-r$
color--magnitude diagram in their early-type galaxy SDSS sample (their
Figure 24). We note that one can detect the effect of not accounting
for evolution by comparing the results of Blanton et al. (in
preparation) for the evolution-corrected luminosity function at
$z=0.1$ to the $1/V_{\mathrm{max}}$-based estimate of the
$\band{0.1}{i}$-band luminosity function presented here (our
luminosity function extends to slightly higher luminosities than that
estimate).

The scope of this paper has not allowed us to delve into a more
detailed discussion of the individual relationships and their relation
to galaxy formation theory. However, we could not do justice to these
topics in the space and time available here, and leave this for the
future, encouraging readers to use the data distributed here to
perform similar work. We note that \citet{kauffmann02a} have performed
a similar analysis using fits to stellar population models. They also
see bimodality in the distribution of galaxy properties, particular in
their measure of the 4000 \AA\ break. In addition, they find that
because their estimated mass-to-light ratio (in their case in the
$\band{0.1}{z}$-band) is a strong (and increasing) function of mass or
luminosity, the dependence of surface mass density in a galaxy on
total stellar mass is monotonic, even though the dependence of surface
brightness on luminosity shows a maximum surface brightness at around
an absolute magnitude of $M_\ast$. A particularly interesting
application along these lines would be to consider the dependence of
local density on stellar mass. 

Two important photometric parameters missing from this space are color
gradients and axis ratios. As it turns out, the color gradients of
galaxies in our sample are usually quite small, partly because the
central bulges of our spiral galaxies are blurred with the disks by
seeing. Nevertheless, using a restricted sample of nearby galaxies,
one can characterize this distribution. The distribution of axis
ratios is related to the three-dimensional shape of the galaxy, so
understanding the conditional distribution of this quantity on other
parameters will yield a better understanding of galaxy
geometry. Preliminary tests have shown that the axis ratios of our
red, concentrated objects tend to be close to unity, whereas the axis
ratios of blue, unconcentrated objects tend to differ significantly
from unity, agreeing with the common understanding of the nature of
elliptical and spiral galaxies. Including a characterization of the
two-dimensional shape of each galaxy would be an interesting way to
expand the results found here.

We began this paper by noting that many galaxy properties have been
known for decades to be correlated, and in particular to be correlated
with position on the Hubble Sequence. Many readers will ask how
galaxies of different morpholopgical classifications are distributed
in this space, a question we could address using a nearby subsample of
the galaxies studied here. We have not done so here because we believe
that morphological classification is not a sufficiently specified
measurement to be straightforwardly interpreted. The position along
the Hubble Sequence is determined by most galaxy classifiers from a
consideration of a galaxy's surface brightness, smoothness,
concentration, axis ratio, the prominence of dust lanes, and spiral
arm pitch angle. However, astronomy has not standardized the weights
to be accorded by the classifier to each of these qualities of a
galaxy image when placing each galaxy along the one-dimensional Hubble
Sequence. In addition, the decision about a galaxy's classification is
highly dependent on the observing conditions, especially the distance
of the galaxy from the observer, the dynamic range of the image, the
passband of the observation, and the angle from which the galaxy is
observed. Even when considering a single image and allowing only the
classifier to vary, the repeatability of classification appears to be
low ($\sigma \sim 2$ in units of Revised Hubble $T$ type;
\citealt{naim95a}). Even this level of repeatability is not clearly
due to the fact that classifiers all weight the various Hubble type
criteria similarly, since the properties under consideration all
correlate.

Nevertheless, as reviewed by \citet{roberts94a}, morphological
classification is clearly important, since it does correlate with many
physical properties of galaxies. Furthermore, we {\it can} specify
certain aspects which determine morphological classification. For
example, many investigators have quantified measures of surface
brightness, concentration, smoothness/blobbiness
(e.g. \citealt{naim97a}), and lopsidedness
(e.g. \citealt{rudnick00a}), among others, and investigated the
dependence of these measures on observing conditions. These efforts,
of which this paper is a part, are forming a new approach to
quantitative morphology that we hope will have a more direct and
better specified connection to theoretical predictions.

\acknowledgments

MB and DWH acknowledge NASA NAG5-11669 for partial support.
MB is grateful for the hospitality of the Department of Physics and
Astronomy at the State University of New York at Stony Brook, who
kindly provided computing facilities on his frequent visits there.

Funding for the creation and distribution of the SDSS has been
provided by the Alfred P. Sloan Foundation, the Participating
Institutions, the National Aeronautics and Space Administration, the
National Science Foundation, the U.S. Department of Energy, the
Japanese Monbukagakusho, and the Max Planck Society. The SDSS Web site
is {\tt http://www.sdss.org/}.

The SDSS is managed by the Astrophysical Research Consortium (ARC) for
the Participating Institutions. The Participating Institutions are The
University of Chicago, Fermilab, the Institute for Advanced Study, the
Japan Participation Group, The Johns Hopkins University, Los Alamos
National Laboratory, the Max-Planck-Institute for Astronomy (MPIA),
the Max-Planck-Institute for Astrophysics (MPA), New Mexico State
University, University of Pittsburgh, Princeton University, the United
States Naval Observatory, and the University of Washington.

\newpage

\clearpage
\clearpage

\setcounter{thefigs}{0}

\clearpage
\stepcounter{thefigs}

\clearpage
\stepcounter{thefigs}

\clearpage
\stepcounter{thefigs}

\clearpage
\stepcounter{thefigs}

\clearpage
\stepcounter{thefigs}

\clearpage
\stepcounter{thefigs}

\clearpage
\stepcounter{thefigs}

\clearpage
\stepcounter{thefigs}

\clearpage
\stepcounter{thefigs}

\clearpage
\stepcounter{thefigs}

%

\clearpage
\stepcounter{thefigs}

\clearpage
\stepcounter{thefigs}

\clearpage
\stepcounter{thefigs}

\clearpage
\stepcounter{thefigs}



\clearpage
 
\begin{deluxetable}{cccrrrrr}
\tablewidth{0pt}
\tablecolumns{8}
\tablecaption{\label{cmrtable} Fits to Median Galaxy Properties as a Function
of Luminosity} 
\tablehead{ Galaxy Type & Property & Units & $\sigma$ & $p_0$ & $p_1$ & $p_2$ & $p_3$}
\startdata
      Concentrated ($n>3.0$) &                                $\band{0.1}{(u-g)}$ &                           mags &  0.321 &  1.697 &-0.0338 & 0.0023 & -0.01005 \cr
                               &                                $\band{0.1}{(g-r)}$ &                           mags &  0.068 &  0.937 &-0.0238 &-0.0039 & -0.00400 \cr
                               &                                $\band{0.1}{(r-i)}$ &                           mags &  0.036 &  0.410 &-0.0082 &-0.0007 & -0.00077 \cr
                               &                                $\band{0.1}{(i-z)}$ &                           mags &  0.076 &  0.322 & 0.0132 &-0.0132 & -0.00053 \cr
                               &                              $\mu_{\band{0.1}{i}}$ &         mags in 1 arcsec${^2}$ &  0.542 & 19.131 &-0.1866 & 0.0759 &  0.00981 \cr
                               &                                   $\log_{10} \rho$ &                            --- &  0.344 &  0.373 &-0.0413 & 0.0168 & -0.00269 \cr
                               &    $\log_{10} [r_{i,50}/ (1~h^{-1}\mathrm{~kpc})]$ &                            --- &  0.109 &  0.305 &-0.2370 & 0.0154 &  0.00277 \cr
         Exponential ($n<1.5$) &                                $\band{0.1}{(u-g)}$ &                           mags &  0.262 &  1.142 &-0.0712 & 0.0095 & -0.00076 \cr
                               &                                $\band{0.1}{(g-r)}$ &                           mags &  0.143 &  0.621 &-0.0606 &-0.0045 &  0.00175 \cr
                               &                                $\band{0.1}{(r-i)}$ &                           mags &  0.063 &  0.352 &-0.0205 &-0.0006 &  0.00013 \cr
                               &                                $\band{0.1}{(i-z)}$ &                           mags &  0.132 &  0.210 &-0.0139 &-0.0194 &  0.00243 \cr
                               &                              $\mu_{\band{0.1}{i}}$ &         mags in 1 arcsec${^2}$ &  0.600 & 20.266 & 0.2545 & 0.0348 & -0.00230 \cr
                               &                                   $\log_{10} \rho$ &                            --- &  0.371 &  0.283 & 0.0080 &-0.0298 &  0.00782 \cr
                               &    $\log_{10} [r_{i,50}/ (1~h^{-1}\mathrm{~kpc})]$ &                            --- &  0.123 &  0.533 &-0.1468 & 0.0061 & -0.00030 \cr
\enddata
\tablecomments{Lists fit parameters $p_i$ (see Equation
\ref{median_fit} in the text) for galaxy properties as a function of
absolute magnitude $M_{\band{0.1}{i}}$, for our exponential and de
Vaucouleurs galaxies separately. One can use these relations to
predict the galaxy properties given the luminosity, but not
vice-versa.}
\end{deluxetable}

\end{document}